\documentclass[lettersize,journal]{IEEEtran}
\usepackage{amsmath,amsfonts}
\usepackage{xcolor}
\usepackage{algorithm}
\usepackage{array}
\usepackage{textcomp}
\usepackage{url}
\usepackage[hidelinks]{hyperref}
\usepackage{verbatim}
\usepackage{graphicx}
\usepackage{cite}
\usepackage{stfloats}
\usepackage{cuted}
\usepackage{amssymb}
\usepackage{algpseudocode}
\usepackage{float}
\usepackage{pdfpages}
\usepackage{svg}
\usepackage{mathrsfs}
\usepackage{subcaption}
\usepackage[justification=raggedright,singlelinecheck=false]{caption}
\def\BibTeX{{\rm B\kern-.05em{\sc i\kern-.025em b}\kern-.08em
    T\kern-.1667em\lower.7ex\hbox{E}\kern-.125emX}}
\usepackage{balance}

\begin{document}

\title{Cooperative ISAC for LAE: Joint Trajectory Planning,\\Power Allocation, and Dynamic Time Division}

\author{
    Fangzhi~Li,
    Zhichu~Ren,
    Cunhua~Pan,~\IEEEmembership{Senior~Member,~IEEE},
    Hong~Ren,~\IEEEmembership{Member,~IEEE},\\
    Jing~Jin,
    Qixing~Wang,
    and~Jiangzhou~Wang,~\IEEEmembership{Fellow,~IEEE}

\thanks{F. Li, Z. Ren, C. Pan, H. Ren, and J. Wang are with National Mobile Communications Research Laboratory, Southeast University, Nanjing, China. J. Jin and Q. Wang are with the China Mobile Research Institute, Beijing, China.}}


\maketitle

\begin{abstract}
To enhance the performance of aerial-ground networks, this paper proposes an integrated sensing and communication (ISAC) framework for multi-UAV systems. In our model, ground base stations (BSs) cooperatively serve multiple unmanned aerial vehicles (UAVs), employing a dynamic time-division strategy where beam scanning for sensing precedes data communication in each time slot. To maximize the sum communication rate while satisfying a mission-level cumulative radar mutual information (MI) requirement, we jointly optimize the UAV trajectories, communication and sensing power allocation, and the time-division ratio. The resulting highly coupled non-convex optimization problem is efficiently solved using an alternating optimization (AO) and successive convex approximation (SCA) framework, which yields a non-decreasing objective sequence and convergence to a finite objective value under the adopted surrogate-based iterative procedure. Extensive simulation results demonstrate that our proposed joint design significantly outperforms benchmark schemes with static trajectories, partially optimized resources, or non-cooperative single-BS transmission. Furthermore, a comprehensive sensitivity analysis reveals the distinct mechanisms by which sensing thresholds and the number of UAVs influence resource allocation and spatial organization, highlighting the critical importance of dynamic, multi-dimensional resource management for effectively navigating the sensing-communication trade-off in low-altitude economies.
\end{abstract}


\section{Introduction}

\IEEEPARstart{T}{he} low-altitude economy (LAE) increasingly relies on fleets of unmanned aerial vehicles (UAVs) that require high-rate communication and timely situational awareness along dynamic flight paths. Traditional separation of sensing and communication suffers from limited spectrum efficiency and weak adaptability to rapid geometric changes in aerial operations \cite{uav_comm_survey_1,uav_comm_survey_2}. Integrated sensing and communication (ISAC) provides a unified framework in which a shared radio stack and common resources support both functions, allowing real-time directional sensing to assist beam alignment and scheduling \cite{survey_isac_1,survey_isac_2,survey_isac_3,10906066}. In particular, cooperative ISAC, where multiple base stations (BSs) jointly serve UAVs, improves both communication and sensing. By exploiting spatial diversity and coordinated transmission, it enhances link reliability and data rates while improving sensing accuracy through multi-view fusion and joint parameter estimation \cite{cheng2024networked,zhang2024joint}.

Recent LAE-oriented studies further highlight the value of infrastructure cooperation, where multiple BSs coordinate to stabilize coverage and reduce reacquisition time for maneuvering UAVs. Cheng et al. \cite{cheng2025networked} studied coordinated transmit beamforming and UAV trajectory design in networked ISAC systems and showed clear gains in communication throughput and sensing reliability. Pan et al. \cite{pan2023cooperative} also developed a cooperative trajectory planning and resource allocation framework that balances sensing and communication under energy and mobility constraints.

A slot-synchronous design with a short sensing preamble followed by data transmission is well suited to mobility-dominated environments. As UAVs maneuver, directional uncertainty accumulates, and brief scans that aggregate echoes to update angular signatures can improve subsequent link formation and interference management \cite{uav_isac_survey_1,uav_isac_survey_2}. Existing ISAC studies have demonstrated the value of sharing time and power between sensing and communication, including periodic-scan designs and joint resource optimization under different mission settings \cite{9858656,meng2022uav,uav_isac_joint_2}. For example, Meng et al. \cite{9858656} proposed a throughput maximization framework for UAV-enabled integrated periodic sensing and communication, highlighting the role of temporal resource allocation. Meanwhile, trajectory control helps maintain favorable geometry, preserve line-of-sight (LoS) conditions, and reduce path loss, thereby easing sensing requirements and improving data rates \cite{uav_traj_comm_1,uav_comm_survey_2}. Building on this idea, recent work has coupled UAV maneuvering with transmission design in ISAC systems and reported gains in energy efficiency, fairness, and localization accuracy \cite{uav_isac_joint_1,uav_isac_joint_3,jing2024isac,pan2023cooperative}. Beyond single-link settings, infrastructure cooperation through coordinated beamforming and trajectory design exploits multiple vantage points to reduce blind sectors and share updated signature information across vehicles \cite{wang2024isac,li2026cooperative,10906066,zhang2024joint}.

Complementing these system designs, information-theoretic and estimation-theoretic perspectives provide fundamental insight into sensing quality quantification and temporal resource management \cite{resource_alloc_isac_1,tradeoff_isac_1,tradeoff_isac_2}. Mission-level constraints based on cumulative mutual information (MI) capture how multiple short scans jointly build evidence for reliable beam alignment, while Cram\'er--Rao bound analyses reveal how power, bandwidth, and aperture size affect angular precision \cite{sensing_metric_mi,sensing_metric_crb,sensing_metric_tradeoff}. Keskin et al. \cite{tradeoff_isac_1} investigated fundamental trade-offs in monostatic ISAC and emphasized the interplay between communication and sensing under resource constraints. Related advances in target localization using standardized multicarrier waveforms also support cooperative sensing and data fusion \cite{zhang2024joint}. Meanwhile, dynamic slot adjustment and predictive scheduling highlight the importance of slot-level responsiveness when motion patterns, queue dynamics, and backhaul loads vary on comparable timescales \cite{pang2024dynamic,khalili2024efficient,zhou2024temporal,chai2024precoding}.

Despite these advances, several challenges remain in achieving end-to-end efficiency for cooperative multi-UAV ISAC systems under mobility. The time division between sensing and communication is often predetermined or only coarsely adjusted, rather than jointly optimized with trajectories and power allocation, even though it directly governs the communication--sensing trade-off in rapidly varying geometries \cite{9858656,meng2022uav,uav_isac_joint_2,pang2024dynamic}. Sensing performance is also often characterized by instantaneous metrics instead of accumulated information budgets that better capture how multiple short preambles support subsequent transmissions. The use of mission-level MI constraints remains limited in networked multi-UAV settings \cite{sensing_metric_mi,resource_alloc_isac_1,tradeoff_isac_1}. In addition, infrastructure cooperation is typically coordinated on slower timescales than per-slot scheduling or studied under only partially optimized resources, which limits its ability to reduce directional uncertainty when mobility and communication backlogs require fast adaptation \cite{cheng2024networked,cheng2025networked,wang2024isac,pan2023cooperative}. Moreover, the resulting optimization couples spatial, temporal, and energy variables in a highly nonconvex manner; although alternating optimization and majorization--minimization are well-established tools, scalable solvers with convergence guarantees for this specific coupling in cooperative LAE scenarios remain insufficiently explored \cite{sca_tutorial,resource_alloc_isac_2,tradeoff_isac_1,gao2024trajectory}.

Motivated by the above, this paper studies a cooperative ISAC framework in which multiple BSs serve multiple UAVs, and each slot consists of a sensing preamble for beam scanning followed by a communication phase. To capture UAV-induced geometry variation and directional uncertainty, we optimize an adaptive slot-level time-division ratio. To reflect the progressive nature of sequential beam scanning, the sensing requirement is modeled by a cumulative MI constraint over the mission horizon. The objective is to maximize the mission sum communication rate by jointly optimizing UAV trajectories, communication and sensing power allocation, and the dynamic time-division ratio, subject to sensing, mobility, and BS power constraints. The resulting nonconvex problem is solved by an alternating optimization (AO) framework, where each subproblem is handled by successive convex approximation (SCA), yielding a non-decreasing objective sequence and convergence to a finite objective value under the adopted surrogate-based iterative procedure \cite{sca_tutorial}. Numerical results show that the proposed design consistently outperforms fixed-split and decoupled baselines under different load conditions, power budgets, and sensing requirements \cite{cheng2025networked,pan2023cooperative,gao2024trajectory}. The main contributions are summarized as follows:

\begin{itemize}
\item We propose a joint optimization framework for cooperative multi-UAV ISAC networks, where beam scanning acquires angular information for the subsequent communication phase. The framework jointly designs UAV trajectories, communication and sensing power allocation, and the dynamic time-division schedule for adaptive resource management in time-varying air-to-ground environments.

\item We model the sensing requirement by a mission-level cumulative MI constraint that matches the sequential beam-scanning operation. We also clarify that MI serves as an information-theoretic proxy for sensing quality, rather than a direct substitute for a specific detection probability or estimation mean squared error (MSE).

\item We develop an alternating-optimization-based algorithm for the resulting non-convex problem by decomposing it into tractable subproblems, yielding an efficient solution for trajectory, power allocation, and time-division updates.

\item Extensive simulations show that the proposed design consistently outperforms the benchmark schemes, highlighting the benefits of dynamic trajectory control and adaptive resource allocation in balancing sensing and communication.
\end{itemize}

The rest of this paper is organized as follows. Section II introduces the system model and problem formulation. Section III presents the proposed alternating optimization algorithm. Section IV provides numerical results to evaluate the performance of our design, and Section V concludes the paper.

\textit{Notations:} Scalars are denoted by italic letters. Vectors and matrices are denoted by boldface lowercase and uppercase letters, respectively. Calligraphic letters denote sets. $\mathbb{R}^{M \times N}$ and $\mathbb{C}^{M \times N}$ denote the spaces of $M \times N$ real and complex matrices. For a matrix $\mathbf{A}$, $\mathbf{A}^T$ and $\mathbf{A}^H$ denote its transpose and Hermitian transpose. For a vector $\mathbf{a}$, $\|\mathbf{a}\|$ denotes its Euclidean norm. The operator $\otimes$ denotes the Kronecker product.

\section{System Model}
\begin{figure}[h]
    \centering
    \includegraphics[width=0.45\textwidth]{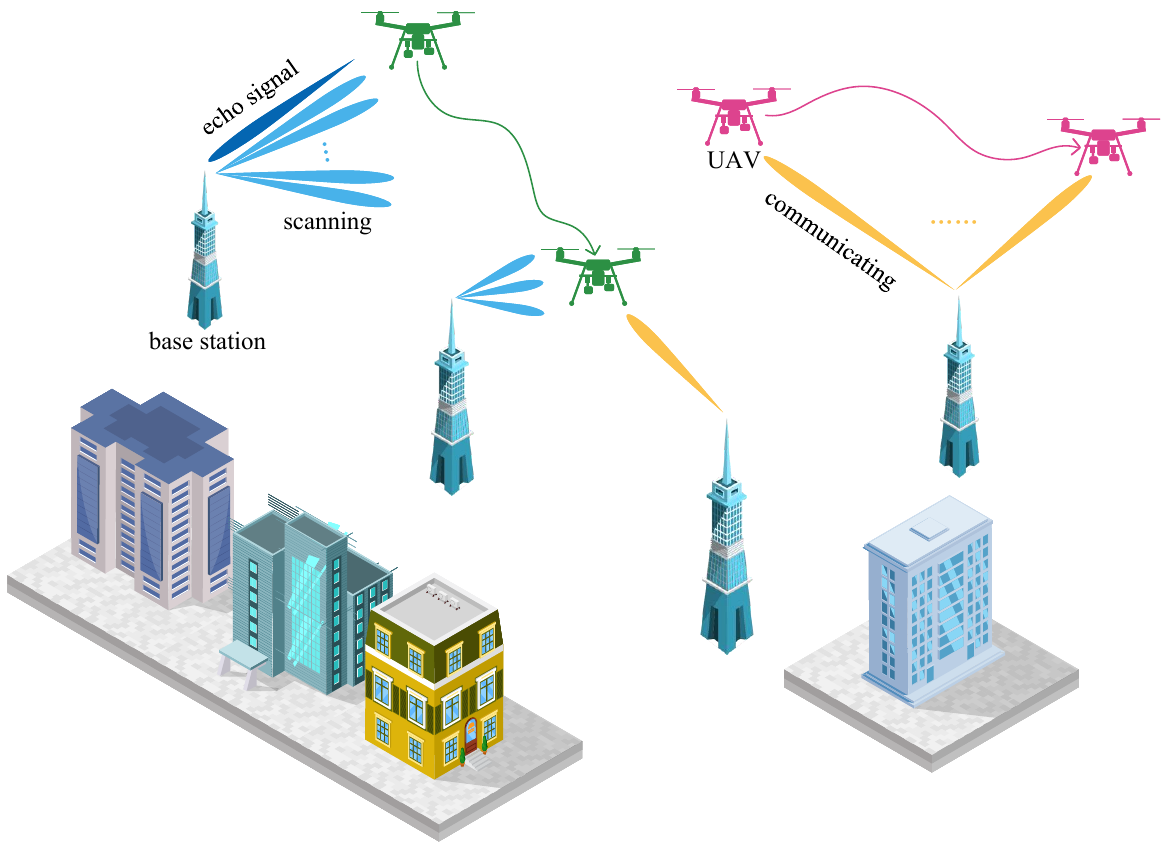}
    \caption{System model.}
    \label{fig:system_model}
\end{figure}

Consider a multi-UAV assisted integrated sensing and communication system, as illustrated in Fig.~\ref{fig:system_model}. The system comprises $K$ UAVs and $M$ ground BSs, denoted by the sets $\mathcal{K} = \{1, 2, \cdots, K\}$ and $\mathcal{M} = \{1, 2, \cdots, M\}$, respectively. The total mission duration is $T$, which is discretized into $N$ equal time slots indexed by $n \in \mathcal{N} = \{1, 2, \cdots, N\}$. The two-dimensional horizontal position of UAV $k \in \mathcal{K}$ at time slot $n$ is represented as $\mathbf{q}_k[n] = (x_k[n], y_k[n]) \in \mathbb{R}^2$, while its flight altitude is assumed to be a constant $H_k$, subject to the constraint $H_{\min} < H_k < H_{\max}$. The general system model allows different UAVs to have different constant altitudes. The optimization formulation remains valid in this general case. For the explicit trajectory-gradient derivation in Section III-C and for all numerical results, we specialize to the equal-altitude case $H_k = H,\forall k$ for notational simplicity. The unequal-altitude case can be handled by the same chain-rule procedure with $H_k$ retained in the corresponding derivative expressions. Each BS $m \in \mathcal{M}$ is located at a fixed horizontal position $\mathbf{v}_m = (v_{m,x}, v_{m,y}) \in \mathbb{R}^2$.

The system operates under a time-division duplexing (TDD) protocol and adopts a quasi-static channel model, where the channel is assumed constant within each time slot of duration $\tau=T/N$. If the channel varies faster due to mobility or tracking errors, the slot duration can be reduced or the design can be updated in a receding-horizon manner using pilots. Such variations mainly reduce beamforming gain and increase residual interference, but the proposed time-division operation and AO--SCA framework remain applicable.

\begin{figure}[h]
    \centering
    \includegraphics[width=0.45\textwidth]{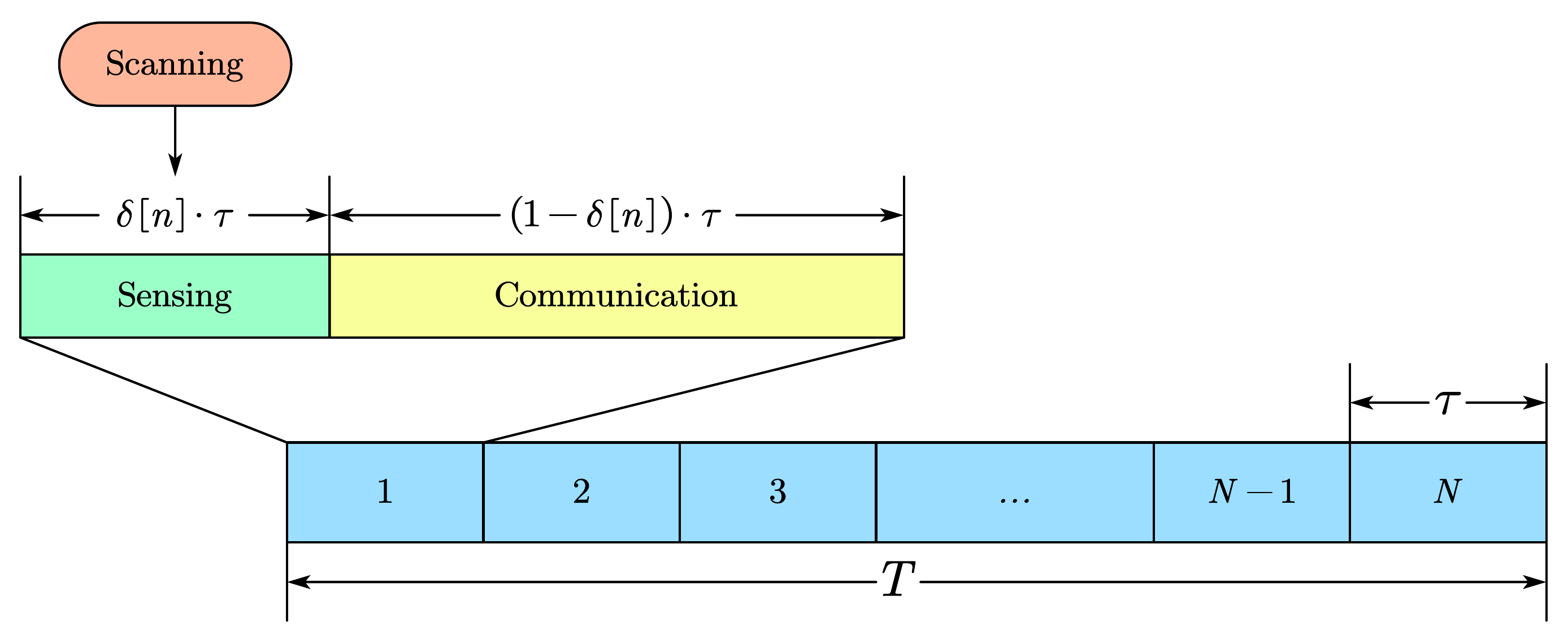}
    \caption{Time slot division.}
    \label{fig:time_slot}
\end{figure}

Each time slot is partitioned into two non-overlapping phases: a sensing phase followed by a communication phase, as shown in Fig.~\ref{fig:time_slot}.

In Phase I, the sensing phase, all BSs perform directional beam scanning to sense UAV positions and angular features. This phase lasts $\delta[n]\tau$, where $0\le \delta[n]\le 1$ for all $n\in\mathcal N$. BS $m$ uses a predefined two-dimensional beamforming codebook $\mathcal{B}_m=\{\mathbf{w}^m_{q,l}\}$, where $\mathbf{w}^m_{q,l}$ steers toward zenith angle $\theta_q$ and azimuth angle $\phi_l$. The zenith range $[\theta_{\min},\theta_{\max}]$ is uniformly quantized into $Q$ directions and the azimuth range $[\phi_{\min},\phi_{\max}]$ is quantized into $L$ directions, yielding $QL$ beam directions. Phase I is divided into $QL$ equal-length sub-slots with duration $\tau^{s}_{q,l}=\delta[n]\tau/(QL)$. The BS sequentially transmits probing beams and receives echoes to estimate coarse UAV positions and angular signatures, which are then used in Phase II.

In Phase II, the communication phase, the BSs use the sensed angular information for directional downlink beamforming over the remaining $(1-\delta[n])\tau$ duration. All UAVs are served concurrently, and better sensing in Phase I improves the beamforming quality and throughput in Phase II.

This slot-level design reflects mobility-driven geometry variation. Since BS--UAV geometry and angular uncertainty vary across slots, the desirable sensing--communication split is generally time-varying. We therefore optimize $\delta[n]$ per slot. In implementation, $\delta[n]$ corresponds to allocating the first $N_{\mathrm{s}}[n]$ orthogonal frequency-division multiplexing (OFDM) symbols in each slot to beam scanning and the remaining symbols to downlink transmission, with the beam-sweeping order coordinated among cooperating BSs.

\subsection{Communication Model}
We first adopt a LoS-dominant air-to-ground channel model to highlight the fundamental coupling among trajectory, beam scanning, and time division in mobility-dominated LAE scenarios. Under this setting, the channel is mainly determined by large-scale geometry. The channel power gain from BS $m$ to UAV $k$ at time slot $n$ is
\begin{align}
\beta_{m,k}[n] &= \beta_0 d_{m,k}^{-2}[n]
= \frac{\beta_0}{H_k^2+\|\mathbf{q}_k[n]-\mathbf{v}_m\|^2},
\end{align}
where $\beta_0$ denotes the channel power gain at a reference distance of 1 meter, and $d_{m,k}[n]= \sqrt{H_k^2+\|\mathbf{q}_k[n]-\mathbf{v}_m\|^2}$ denotes the distance from the BS $m$ to UAV $k$. The above LoS model can be extended by incorporating a path-loss exponent, log-normal shadowing, and Rician small-scale fading with an elevation-angle-dependent $K$-factor. In this case, the instantaneous signal-to-interference-plus-noise ratio (SINR) becomes random and the communication objective can be formulated using an ergodic rate, an outage-constrained rate, or a sample-average rate based on channel realizations. The proposed AO--SCA framework remains applicable after constructing convex surrogates for the adopted rate metric.

Each BS is equipped with a uniform planar array (UPA) with $N_{p} = N_{px}N_{py}$ antennas for transmitting and receiving signals. The transmit array response vector of BS $m$ towards UAV $k$ is
\begin{align}
    \mathbf{a}_{m,k}(\mathbf{q}_{k}[n],\mathbf{v}_{m})\!\!= &\mathbf{a}_{m,k,x}(\mathbf{q}_{k}[n],\mathbf{v}_{m}) \otimes \mathbf{a}_{m,k,y}(\mathbf{q}_{k}[n],\mathbf{v}_{m}) \nonumber\\
    =&[1,\cdots,e^{-\frac{\jmath2\pi(N_{px}-1)d_{x}\Phi(\mathbf{q}_{k}[n],\mathbf{v}_{m})}{\lambda}}]^T
    \nonumber \\ 
    & \otimes [1,\cdots,e^{-\frac{\jmath2\pi(N_{py}-1)d_{y}\Omega(\mathbf{q}_{k}[n],\mathbf{v}_{m})}{\lambda}}]^T,
\end{align}
where $\Phi(\mathbf{q}_{k}[n],\mathbf{v}_{m}) = \sin(\theta(\mathbf{q}_{k}[n],\mathbf{v}_{m}))\cos(\phi(\mathbf{q}_{k}[n],\mathbf{v}_{m}))$ and $\Omega(\mathbf{q}_{k}[n],\mathbf{v}_{m}) = \sin(\theta(\mathbf{q}_{k}[n],\mathbf{v}_{m}))\sin(\phi(\mathbf{q}_{k}[n],\mathbf{v}_{m}))$ denote the direction cosines along the $x$ and $y$ axes. $\theta(\mathbf{q}_{k}[n],\mathbf{v}_{m})$ represents the zenith angle of departure (AoD) of the signal from BS $m$ to UAV $k$, and $\phi(\mathbf{q}_{k}[n],\mathbf{v}_{m})$ denotes the azimuth angle. Therefore, the baseband equivalent channel $\mathbf{h}^{c}_{m,k}( \mathbf{q}_{k}[n], \mathbf{v}_{m})$ from BS $m$ to UAV $k$ is
\begin{align}
    \big(\mathbf{h}_{m,k}^{c}(\mathbf{q}_{k}[n],\mathbf{v}_{m}) \big)^{H}
= \sqrt{\beta_{m,k}[n]}\,\mathbf{a}_{m,k}^H(\mathbf{q}_{k}[n],\mathbf{v}_{m}).
\end{align}

Let $s^{c}_{k}[n]$ denote the unit-power communication symbol intended for UAV $k$ at time slot $n$, with $\mathbb{E}[|s^{c}_{k}[n]|^2]=1$. Let $\mathbf{w}^{c}_{m,k}[n]$ denote the normalized beamforming vector at BS $m$ for UAV $k$, satisfying $\|\mathbf{w}^{c}_{m,k}[n]\|^2 = 1$, and let $\eta^{c}_{m,k}[n] \ge 0$ denote the corresponding power allocation coefficient. The transmitted communication signal from BS $m$ for UAV $k$ is
\begin{align}
    \mathbf{x}^{c}_{m,k}[n] = \sqrt{\eta^{c}_{m,k}[n]}\,\mathbf{w}^{c}_{m,k}[n]\,s^{c}_{k}[n].
\end{align}

Accordingly, the received signal at UAV $k$ is
\begin{align}
    y^{c}_{k}[n]
    = &\sum_{m=1}^{M}\big(\mathbf{h}_{m,k}^{c}(\mathbf{q}_{k}[n],\mathbf{v}_{m})\big)^{H}\mathbf{x}^{c}_{m,k}[n] \nonumber\\
    &+ \sum_{m=1}^{M}\sum_{i \in \mathcal{K}\backslash\{k\}}
    \big(\mathbf{h}_{m,k}^{c}(\mathbf{q}_{k}[n],\mathbf{v}_{m})\big)^{H}\mathbf{x}^{c}_{m,i}[n]
    + n_k[n],
\end{align}
where $n_k[n] \sim \mathcal{CN}(0,\sigma_k^2)$ is the additive white Gaussian noise.

We consider noncoherent cooperative multi-BS transmission, where multiple BSs transmit the same user data stream to UAV $k$ without phase synchronization. Hence, the useful received powers from different BSs add noncoherently at the UAV, and the resulting SINR is
\begin{align}
    \gamma^{c}_{k}[n]
    \!=\!
    \frac{\displaystyle \sum_{m \in \mathcal{M}} \!\!\eta^{c}_{m,k}[n]
    \left| \big( \mathbf{h}_{m,k}^{c}(\mathbf{q}_{k}[n],\! \mathbf{v}_{m}) \big)^{H}
    \!\mathbf{w}^{c}_{m,k}[n]\right|^2}
    {\displaystyle \sum_{m \in \mathcal{M}} \sum_{i \in \mathcal{K}\backslash\{k\}}
    \!\!\!\!\eta^{c}_{m,i}[n]
    \left| \big( \mathbf{h}_{m,k}^{c}(\mathbf{q}_{k}[n],\! \mathbf{v}_{m}) \big)^{H}
    \!\mathbf{w}^{c}_{m,i}[n]\right|^2 \!\!\!+\! \sigma_k^2 }.
\end{align}

Then the sum communication rate for all the UAVs can be expressed as
\begin{align} R[n] = & \sum_{k \in \mathcal{K}}(1-\delta[n]) R_{k}[n] \nonumber \\ = & \sum_{k \in \mathcal{K}}(1-\delta[n]) \log_{2}(1 + \gamma^{c}_{k}[n]). \end{align}

\subsection{Sensing Model}
The sensing operation mainly takes place in Phase I. BS $m$ sweeps the beams in the codebook $\mathcal{B}_m = \{\mathbf{w}^m_{1,1},\cdots,\mathbf{w}^m_{Q,1},\mathbf{w}^m_{1,2},\cdots,\mathbf{w}^m_{Q,2},\cdots,\mathbf{w}^m_{1,L},\cdots,\mathbf{w}^m_{Q,L}\}$, where $\mathbf{w}^m_{q,l} = \mathbf{a}_{m}(\theta_{q},\phi_{l})/\Vert\mathbf{a}_{m}(\theta_{q},\phi_{l})\Vert$ with $1 \le q \le Q$ and $1 \le l \le L$. Each BS collects the echoes reflected from the UAVs. At the end of scanning, each BS estimates the target direction and constructs a unit-norm matched filter aligned with the estimated arrival direction for subsequent echo processing.

The direct sensing channel from BS $j$ to BS $m$ is denoted by $\mathbf{G}_{j,m}$, which captures the direct BS-to-BS coupling in Phase I. Since the BS locations are fixed, $\mathbf{G}_{j,m}$ varies slowly and can be tracked by periodic calibration and pilot transmissions. The sensing channel gain including the path loss from BS $j$ to BS $m$ via the reflection of UAV $k$ is represented by $\beta^{s}_{j,m,k}$. Let $\xi_{j,m,k} \sim \mathcal{CN}(0, 1)$ denote the normalized radar cross section (RCS) of the target through the reflection path from BS $j$ to BS $m$. The signal received at BS $m$ in the $q$-th sub-slot of time slot $n$ and reflected by UAV $k$ is given by
\begin{align}
    \mathbf{y}^{s}_{m,k,q,l}[n] \!= \!& \; \xi_{m,m,k}\!\sqrt{\!\beta^{s}_{m,m,k}}\mathbf{a}_{m,k}(\bar{\theta}^{r}_{q},\! \bar{\phi}^{r}_{l})\mathbf{a}^{H}_{m,k}(\bar{\theta}^{t}_{q},\! \bar{\phi}^{t}_{l})\mathbf{x}_{m,q,l}[n] \nonumber\\
    & + \sum_{j\in \mathcal{M} \backslash \{m\}} \mathbf{G}_{j,m} \mathbf{x}_{j,q,l}[n] \nonumber\\
    & + \sum_{j\in \mathcal{M} \backslash \{m\}} \xi_{j,m,k}\sqrt{ \beta^{s}_{j,m,k}}\,\mathbf{a}_{m,k}(\bar{\theta}^{r}_{q},\bar{\phi}^{r}_{l}) \nonumber \\ & \qquad \cdot \mathbf{a}^{H}_{j,k}(\bar{\theta}^{t}_{q},\bar{\phi}^{t}_{l})\mathbf{x}_{j,q,l}[n] + \mathbf{n}_{m,q,l}[n],
\end{align}
where $\bar{\phi}^{r}_{l}$ and $\bar{\theta}^{r}_{q}$ denote the azimuth and zenith from the true UAV position to the sensing receiver, while $\bar{\phi}^{t}_{l}$ and $\bar{\theta}^{t}_{q}$ denote the azimuth and zenith from the UAV to the transmitter. The transmit signal $\mathbf{x}_{m,q,l}[n]$ is denoted as
\begin{align}
    \mathbf{x}_{m,q,l}[n]=\sqrt{\eta^{s}_{m,q,l}[n]}\,\mathbf{w}_{q,l}^{m}\,s_{m,q,l}[n],
\end{align}
where $\eta^{s}_{m,q,l}[n]$ is the power allocation coefficient at BS $m$, $s_{m,q,l}[n]$ is the sensing symbol with $\mathbb{E}[|s_{m,q,l}[n]|^{2}]=1$, and $\mathbf{n}_{m,q,l}[n]\sim\mathcal{CN}(\mathbf{0},\sigma_m^{2}\mathbf{I})$ denotes the receiver noise.

The direct-link interference can be mitigated through standard calibration and digital cancellation. The BSs periodically transmit orthogonal pilots so that each receiver estimates the slowly varying coupling channel $\mathbf{G}_{j,m}$. Since the sensing waveforms and beam indices in Phase I are coordinated via the backhaul, BS $m$ can reconstruct $\sum_{j\in\mathcal{M}\backslash\{m\}}\hat{\mathbf{G}}_{j,m}\mathbf{x}_{j,q,l}[n]$ and subtract it in digital baseband before matched filtering. The pilot overhead is modest because $\mathbf{G}_{j,m}$ changes slowly over time.

After subtracting the reconstructed direct-link interference, the residual array output at BS $m$ is
\begin{align}
\mathbf{y}^{s}_{m,q,l}[n] \!\!
=& \!\! \sum_{k\in\mathcal{K}} \! \sum_{j\in\mathcal{M}} \!\!
\xi_{j,m,k}\!\sqrt{\!\beta^{s}_{\!j,m,k}}
\mathbf{a}_{m,k}(\bar{\theta}^{r}_{\!q}\!,\! \bar{\phi}^{r}_{l})
\mathbf{a}^{H}_{j,k}(\bar{\theta}^{t}_{\!q}\!,\! \bar{\phi}^{t}_{l})
\mathbf{x}_{\!j,q,l}\![n] \nonumber\\
&\; + \mathbf{n}_{m,q,l}[n].
\label{eq:sensing_signal}
\end{align}

Define the unit-norm combiner from the scan-based angle estimate $(\hat\theta^r_q,\hat\phi^r_l)$, then we have
\begin{align}
    \mathbf{u}_{m,q,l} \triangleq & \frac{\mathbf{a}_{m}(\hat\theta^r_q,\hat\phi^r_l)}{\|\mathbf{a}_{m}(\hat\theta^r_q,\hat\phi^r_l)\|}, \\
    \tilde y^{s}_{m,q,l}[n] \triangleq &\mathbf{u}_{m,q,l}^{H}\,\mathbf{y}^{s}_{m,q,l}[n].
\end{align}

The vector $\mathbf{u}_{m,q,l}$ is a unit-norm receive combiner steered toward the estimated receive angle, and $\tilde y^{s}_{m,q,l}[n]=\mathbf{u}_{m,q,l}^{H}\mathbf{y}^{s}_{m,q,l}[n]$ is the corresponding beamformed scalar observation. Since $\mathbf{u}_{m,q,l}$ has unit norm, the combined noise remains $\tilde n_{m,q,l}[n]\sim\mathcal{CN}(0,\sigma_m^2)$.

After direct-link cancellation and matched filtering, the sensing SINR at BS $m$ is obtained in average-power form by taking expectations over the sensing symbols and normalized RCS, which yields the deterministic quadratic form in \eqref{eq:sensing_SINR}.
\begin{figure*}
    \begin{equation}
        \gamma^{s}_{m,q,l}[n] = 
        \frac{\displaystyle \sum_{k \in \mathcal{K}} \beta^{s}_{m,m,k} \eta^{s}_{m,q,l}[n] \left| \mathbf{u}_{m,q,l}^{H}\, \mathbf{a}_{m,k}(\bar{\theta}^{r}_{q},\bar{\phi}^{r}_{l})\,\mathbf{a}^{H}_{m,k}(\bar{\theta}^{t}_{q},\bar{\phi}^{t}_{l})\,\mathbf{w}_{q,l}^{m}\right|^2 }
        {\displaystyle \sum_{k \in \mathcal{K}}\sum_{j\in \mathcal{M} \backslash \{m\}} \beta^{s}_{j,m,k} \eta^{s}_{j,q,l}[n] \left| \mathbf{u}_{m,q,l}^{H}\, \mathbf{a}_{m,k}(\bar{\theta}^{r}_{q},\bar{\phi}^{r}_{l})\,\mathbf{a}^{H}_{j,k}(\bar{\theta}^{t}_{q},\bar{\phi}^{t}_{l})\,\mathbf{w}_{q,l}^{j}\right|^2 
        + \sigma_{m}^2 }.
    \label{eq:sensing_SINR}
    \end{equation}
\end{figure*}

In \eqref{eq:sensing_SINR}, the reflected components associated with the active sensing operation of BS $m$ are aggregated in the numerator as useful sensing returns, since they correspond to the intended beam-scanning reception at BS $m$ in the current angle bin. By contrast, the reflected components generated by other BSs with $j\neq m$ are treated as interference because they arise from simultaneously transmitted beams that are not coherently processed by receiver $m$.

Based on the sensing SINR, we adopt radar mutual information, abbreviated as MI, as a tractable proxy for sensing quality. Under the matched-filtered scalar Gaussian observation model, $\log_2(1+\gamma^{s}_{m,q,l}[n])$ quantifies the information contributed by one sensing sub-slot. For Gaussian linear models, MI increases monotonically with sensing SINR and is consistent with improved estimation and detection performance. Although MI does not map to a unique estimation error or detection probability, it provides a unified metric for sensing-resource allocation in sequential beam scanning.

By leveraging the sensing SINR, the radar MI for BS $m$ at angle bin $(q,l)$ in time slot $n$ is expressed as
\begin{align}
    R^{\mathrm{rad}}_{m,q,l}[n] = \log_2\left(1 + \gamma^{s}_{m,q,l}[n]\right).
\end{align}

Since beam scanning is repeated over angle bins and time slots, the sensing information accumulates over the mission. We therefore define the cumulative radar MI as
\begin{align}
    R^{\mathrm{rad}} = \sum_{n \in \mathcal{N}} \frac{\delta[n]}{QL} \sum_{m \in \mathcal{M}} \sum^{Q}_{q=1} \sum^{L}_{l=1} R^{\mathrm{rad}}_{m,q,l}[n].
\end{align}

The weighting factor $\delta[n]/(QL)$ reflects the relative sensing time allocated to each beam-scanning sub-slot in time slot $n$. Here the radar-MI metric is normalized with respect to slot duration and bandwidth, so only the slot-level sensing-time fraction appears explicitly. We adopt this cumulative MI metric because the beam-scanning process in different slots contributes complementary information for subsequent beam alignment and communication, and a mission-level sensing requirement is more consistent with such sequential evidence accumulation than a stringent per-slot sensing constraint. In particular, a cumulative constraint allows the scheduler to allocate more sensing effort to geometrically difficult slots and less sensing effort to favorable slots, thereby improving overall communication efficiency while still guaranteeing a target sensing-information budget over the entire mission.

Therefore, we impose the following mission-level sensing requirement:
\begin{align}
    R^{\mathrm{rad}} \ge R_{\min}^{\mathrm{MI}},
\end{align}
where $R_{\min}^{\mathrm{MI}}$ is the required sensing-information budget for downstream sensing-assisted communication operations such as beam alignment and tracking.

\subsection{Practical Considerations and Extensions}

The proposed formulation adopts idealized assumptions to make the core sensing-communication coupling transparent. In practice, imperfect channel state information (CSI), sensing errors, synchronization mismatch, and hardware limitations may reduce the effective communication and sensing SINRs, thereby affecting the optimized time division and power allocation. These effects can be incorporated through robust or outage-aware rate formulations, statistical angular-error models, elevated effective noise/interference terms, or discrete and slowly varying \(\delta[n]\) constraints. Importantly, such modifications mainly affect the coefficients and uncertainty models in the optimization problem, while the proposed AO--SCA framework remains applicable after constructing suitable convex surrogates.

\subsection{Problem Formulation}

In the proposed system, we jointly optimize the communication power allocation $\{\boldsymbol{\eta}^{c}[n]\}$, sensing power allocation $\{\boldsymbol{\eta}^{s}[n]\}$, sensing duration ratio $\{\delta[n]\}$, and UAV trajectories $\{\mathbf{q}_{k}[n]\}$ to maximize the mission sum communication rate subject to the cumulative radar MI requirement, UAV mobility constraints, and BS energy budgets.

Each UAV starts from $\mathbf{q}_{k}^{\mathrm{I}}$ and ends at $\mathbf{q}_{k}^{\mathrm{F}}$:
\begin{align}
    \mathbf{q}_{k}[1] &= \mathbf{q}_{k}^{\mathrm{I}}, \label{eq:init_position_rewrite} \\
    \mathbf{q}_{k}[N] &= \mathbf{q}_{k}^{\mathrm{F}}, \label{eq:final_position_rewrite}
\end{align}

Each UAV also satisfies the maximum-speed constraint and the minimum-separation constraint:
\begin{alignat}{2}
    &\Vert \mathbf{q}_{k}[n\!+\!1] \!-\! \mathbf{q}_{k}[n] \Vert^2 \!\le\! (V_{\max}\tau)^2,
    \ \forall k \!\in\! \mathcal{K},n \!\in\! \{1,\ldots,N\!-\!1\}, \label{eq:speed_constraint_rewrite} \\
    &\Vert \mathbf{q}_{j}[n] \!-\! \mathbf{q}_{i}[n] \Vert^2 \!+\! (H_{j} \!-\! H_{i})^2 \!\ge\! D_{\min}^2,
    \ \forall i,j \!\in\! \mathcal{K},i \!\ne\! j,n \!\in\! \mathcal{N}, \label{eq:collision_avoidance_rewrite}
\end{alignat}
where $V_{\max}$ is the maximum UAV speed and $D_{\min}$ is the minimum safe distance.

Each BS follows a communication power budget $P_{\max}^{c}$ and a per-slot sensing energy budget $W_{\max}^{s}$, where $W_{\max}^{s}=P_{\max}^{s}\tau$. The constraints for BS $m$ at slot $n$ are
\begin{alignat}{2}
    &\sum_{k \in \mathcal{K}} \eta_{m,k}^{c}[n] \le P^{c}_{\max},
    &&\ \forall m,n, \label{eq:c_power_constraint} \\
    &\frac{\delta[n]\tau}{QL} \sum_{q=1}^{Q}\sum_{l=1}^{L} \big\|\sqrt{\eta^{s}_{m,q,l}[n]}\,\mathbf{w}^{m}_{q,l}\big\|^{2} \le W^{s}_{\max},
    &&\ \forall m,n, \label{eq:s_power_constraint}
\end{alignat}

Given the above constraints, the joint optimization problem of communication power allocation, sensing power allocation, time-division ratio, and UAV trajectory can be formulated as follows
\begin{subequations}
    \label{eq:problem1}
    \begin{align}
    (\text{P}1): & \max_{\{\boldsymbol{\eta}^{c}[n],\boldsymbol{\eta}^{s}[n],\delta[n],\mathbf{q}_{k}[n]\}} \sum_{n \in \mathcal{N}} R[n] \label{eq:problem1_a}\\
    \text{s.t.} \quad 
    & R^{\mathrm{rad}} \ge R_{\min}^{\mathrm{MI}}, \label{eq:problem1_b}\\
    & \sum_{k \in \mathcal{K}} \eta_{m,k}^{c}[n] \leq P^{c}_{\max},
    \quad \forall m, n, \label{eq:problem1_c}\\
    & \frac{\delta[n]\tau}{QL} \sum^{Q}_{q=1} \sum^{L}_{l=1} \|\sqrt{\eta^{s}_{m,q,l}[n]}\mathbf{w}_{q,l}^{m}\|^{2} \leq W^{s}_{\max}, \quad \forall m, n, \label{eq:problem1_d}\\
    & 0 \le \delta[n] \le 1, \quad \forall n \in \mathcal{N}, \label{eq:problem1_e}\\
    & \mathbf{q}_{k}[1] = \mathbf{q}_{k}^{\mathrm{I}}, \label{eq:problem1_f}\\
    & \mathbf{q}_{k}[N] = \mathbf{q}_{k}^{\mathrm{F}}, \label{eq:problem1_g}\\
    & \|\mathbf{q}_{k}[n\!\!+\!\!1]\!-\!\mathbf{q}_{k}[n]\|^2 \!\leq \!(V_{\max}\tau)^2, \forall k \!\in \!\mathcal{K}, n \!\in \!\{1,...,N\!\!-\!\!1\}, \label{eq:problem1_h}\\
    & \|\mathbf{q}_{j}[n]-\mathbf{q}_{i}[n]\|^2 + (H_{j}-H_{i})^2 \geq D_{\min}^2, \nonumber\\
    & \quad \forall j, i \in \mathcal{K}, j \neq i, n \in \mathcal{N}. \label{eq:problem1_i}
    \end{align}
\end{subequations}

The objective in \eqref{eq:problem1_a} maximizes the total communication rate $\sum_{n\in\mathcal N} R[n]$ over the mission horizon. Under the adopted TDD protocol, $R[n]$ depends on the sensing duration ratio $\delta[n]$, the communication power allocation $\boldsymbol{\eta}^{c}[n]$, and the UAV trajectories $\{\mathbf{q}_k[n]\}$, while the sensing power allocation $\boldsymbol{\eta}^{s}[n]$ is used to satisfy the radar-MI and sensing-power constraints. Constraint \eqref{eq:problem1_b} imposes a cumulative radar-MI requirement over beam directions and time slots, allowing $\delta[n]$ to adapt to time-varying geometry and uncertainty. Constraints \eqref{eq:problem1_c} and \eqref{eq:problem1_d} limit the BS communication and sensing powers. Constraints \eqref{eq:problem1_f} and \eqref{eq:problem1_g} specify the initial and final UAV positions. Constraint \eqref{eq:problem1_h} enforces the maximum-speed limit, and constraint \eqref{eq:problem1_i} guarantees safe separation among UAVs. Together, these constraints ensure the feasibility and safety of the UAV-assisted ISAC system.

Problem P1 is non-convex due to the logarithmic rate expression, the radar-MI term, and the coupling among trajectory, power allocation, and time division variables.

\section{Proposed Solution}
We solve Problem (P1) via an alternating-optimization framework that iteratively updates the communication power allocation, sensing power allocation, sensing duration ratio, and UAV trajectories. The communication-power, time-division, and trajectory updates are designed to improve or preserve the sum-rate objective. The sensing-power update minimizes the total sensing power subject to the radar-MI and sensing-energy constraints, and it does not decrease the communication objective because $\sum_{n\in\mathcal N}R[n]$ is independent of $\{\boldsymbol{\eta}^{s}[n]\}$ under the adopted TDD model.

\subsection{Power Allocation Optimization}

Given the sensing duration ratio $\delta[n]$ and the UAV trajectory $\mathbf{q}_{k}[n]$, the BS transmit power optimization can be decomposed into two subproblems.

\textbf{1) Communication power allocation subproblem:} This subproblem focuses on optimizing the communication power allocation vector $\boldsymbol{\eta}^{c}$ to maximize the communication throughput. It is formulated as
\begin{subequations}
\label{eq:problem2_1}
\begin{align}
    (\text{P}2.1): & \max_{\{\boldsymbol{\eta}^{c}[n]\}} \sum_{n \in \mathcal{N}} R[n] \label{eq:problem2_1_a}\\
    \text{s.t.} \quad 
    & \eqref{eq:problem1_c}, \nonumber
\end{align}
\end{subequations}
where constraint \eqref{eq:problem1_c} ensures that the total communication power at each BS does not exceed its maximum transmit power.

Constraint \eqref{eq:problem1_c} is affine in $\{\boldsymbol{\eta}^{c}[n]\}$ and hence convex, while the objective $\sum_{n\in\mathcal N}R[n]$ is non-convex. We rewrite each per-UAV rate as a difference of two concave terms and linearize the interference-related term to obtain a tractable surrogate.
\begin{align}
R_k[n]\!\!
= &\log_2\!\!\big(1+\gamma_k^{c}[n]\big) \nonumber\\
= &\log_2\!\!\Big(\!\!\sum_{i\in\mathcal K}\!\sum_{m\in\mathcal M}\!\!
\eta^{c}_{m,i}[n]\!\big|\mathbf h^{c}_{m,k}(\mathbf q_k[n],\! \mathbf v_m)^{\!H}\mathbf{w}^{c}_{m,i}[n]\!\big|^2
\!\!+\!\sigma_k^{2}\!\!\Big)\nonumber \\
&- \hat R_k[n], \label{eq:Rk_diff_final}
\end{align}
where
\begin{align}
\hat R_k[n]\!\!\triangleq\!\!
\log_2\!\!\Big(\!\!\!\sum_{i\in\mathcal K\setminus\{k\}}\!\sum_{m\in\mathcal M}
\!\!\!\eta^{c}_{m,i}\![n]\!\big|\mathbf h^{c}_{m,k}\!(\mathbf q_k\![n],\!\mathbf v_m)^{\!H}\!\mathbf w^{c}_{m,i}\![n]\big|^2
\!\!\!\!+\!\sigma_k^{2}\!\!\Big). \label{eq:Rhat_def_final}
\end{align}

To obtain a tractable surrogate, we linearize $\hat R_k[n]$ with respect to the communication power variables around the current feasible point $\boldsymbol{\eta}^{c(f)}$, while treating the beamformers $\{\mathbf w^{c}_{m,i}[n]\}$ as fixed parameters determined by the current sensing-assisted beam design. The first-order Taylor expansion yields the global upper bound
\begin{align}
\hat{R}_k[n] \!\!
\le &
\hat{R}_k^{(f)}[n] \!\!
+\!\!\! \sum_{i\in\mathcal K\setminus\{k\}}\sum_{m\in\mathcal M}
\!\! B^{c(f)}_{m,i,k}[n]\!\!\left(\eta^{c}_{m,i}[n]\!-\!\eta^{c(f)}_{m,i}[n]\right) \nonumber
\\ & \triangleq \hat R^{cp}_k[n], \label{eq:Rhat_upper_new}
\end{align}
where $\hat{R}_k^{(f)}[n]$ and $B^{c(f)}_{m,i,k}[n]$ are shown in~\eqref{eq:Rhatf_new} and ~\eqref{eq:Bcoef_new}.
\begin{figure*}
\begin{gather}
\hat{R}_k^{(f)}[n]
=\log_2\!\Bigg(
\sum_{i\in\mathcal K\setminus\{k\}}\sum_{m\in\mathcal M}
\eta^{c(f)}_{m,i}[n]\;
\big|\mathbf h^{c}_{m,k}(\mathbf q_k[n],\mathbf v_m)^{H}\mathbf w^{c}_{m,i}[n]\big|^{2}
+\sigma_k^{2}\Bigg). \label{eq:Rhatf_new}
\\
B^{c(f)}_{m,i,k}[n]
=\log_2(e)\;
\frac{\big|\mathbf h^{c}_{m,k}(\mathbf q_k[n],\mathbf v_m)^{H}\mathbf w^{c}_{m,i}[n]\big|^{2}}
{\displaystyle
\sum_{i'\in\mathcal K\setminus\{k\}}\sum_{m'\in\mathcal M}
\eta^{c(f)}_{m',i'}[n]\;
\big|\mathbf h^{c}_{m',k}(\mathbf q_k[n],\mathbf v_{m'})^{H}\mathbf w^{c}_{m',i'}[n]\big|^{2}
+\sigma_k^{2}}.
\label{eq:Bcoef_new}
\end{gather}
\end{figure*}

The bound in \eqref{eq:Rhat_upper_new} is tight at the current iterate, i.e., $\hat{R}_k[n]=\hat{R}_k^{(f)}[n]$ and the gradients match at $\eta^{c}=\eta^{c(f)}$.
Thus, the per-slot objective admits the following concave lower bound:
\begin{align}
\bar R[n]
\triangleq &\sum_{k\in\mathcal K} (1-\delta[n])\nonumber
\\&\cdot 
\Big[\log_2\!\Big(\!\sum_{i\in\mathcal K}\!\sum_{m\in\mathcal M}
\!\!\!\eta^{c}_{m,i}\![n]\!\big|\mathbf h^{c}_{m,k}\!(\mathbf q_k\![n],\! \mathbf v_m)^{\!H}\!\mathbf w^{c}_{m,i}\![n]\!\big|^{2}
\!\!\!+\!\sigma_k^{2}\!\Big) \nonumber
\\&- \hat R^{cp}_k[n]\Big],
\end{align}
which we maximize in each SCA iteration.

Clearly, \( \bar{R}[n] \) is concave because it is a sum of logarithms of positive affine functions and linear terms in the power variables \( \eta^{c}_{m,i} \). Maximizing a concave objective over a convex feasible set is a convex optimization problem, so the problem can be solved efficiently using the CVX package.

\textbf{2) Sensing power allocation subproblem:} Under the adopted TDD protocol, the communication sum-rate objective $\sum_{n\in\mathcal N}R[n]$ is independent of the sensing power allocation variables $\{\boldsymbol{\eta}^{s}[n]\}$. Therefore, instead of treating the sensing-power update as a rate-maximization step, we formulate it as an auxiliary optimization problem that minimizes the total sensing transmit power while satisfying the sensing-related constraints. This auxiliary objective does not alter the primary objective of Problem $(\text{P}1)$, but helps avoid power-redundant feasible solutions.

The sensing power allocation subproblem is formulated as
\begin{subequations}
\label{eq:problem2_2}
\begin{align}
    (\text{P}2.2): & \min_{\{\boldsymbol{\eta}^{s}[n]\}} \sum_{n \in \mathcal{N}}\sum_{m\in\mathcal M}\sum_{q=1}^{Q}\sum_{l=1}^{L}\eta^{s}_{m,q,l}[n] \label{eq:problem2_2_a}\\
    \text{s.t.} \quad 
    & \eqref{eq:problem1_b},\eqref{eq:problem1_d}. \nonumber
\end{align}
\end{subequations}
where constraint \eqref{eq:problem1_b} guarantees a minimum radar MI for reliable sensing, and \eqref{eq:problem1_d} enforces the sensing power budget at each BS. For simplicity, we define $\mathbf{A}_{j,m,k,q,l} = \sqrt{ \beta^{s}_{j,m,k}}\mathbf{a}_{m,k}(\bar{\theta}^{r}_{q},\bar{\phi}^{r}_{l})\mathbf{a}^{H}_{j,k}(\bar{\theta}^{t}_{q},\bar{\phi}^{t}_{l})$.

It is easy to see that constraint \eqref{eq:problem1_d} is a linear inequality with respect to the optimization variable \( \eta^{s}_{m,q,l}[n] \), and therefore, this constraint is convex. However, constraint \eqref{eq:problem1_b} is non-convex, and we handle it using a similar SCA approach.

For the radar MI of BS $m$ at angle bin $(q,l)$ in slot $n$, we rewrite it as a
difference of two concave terms
\begin{align}
&R^{\mathrm{rad}}_{m,q,l}[n]\nonumber\\
= &\log_2 \Bigg(
\sum_{j\in\mathcal M}\sum_{k\in\mathcal K}
\eta^{s}_{j,q,l}[n]
\big\vert \mathbf{u}_{m,q,l}^{H}\mathbf{A}_{j,m,k,q,l}\mathbf{w}^{j}_{q,l}\big\vert^{2}
+\sigma_m^2 \Bigg) \nonumber
\\ &-\hat R^{\mathrm{rad}}_{m,q,l}[n], \label{eq:R_rad_new}
\end{align}
where
\begin{align}
&\hat R^{\mathrm{rad}}_{m,q,l}[n] \nonumber \\
\triangleq &\log_2 \!\!\Bigg(
\!\sum_{j\in\mathcal M\setminus\{m\}}\!\sum_{k\in\mathcal K}
\!\eta^{s}_{j,q,l}[n]
\big\vert\mathbf{u}_{m,q,l}^{H}\mathbf{A}_{j,m,k,q,l}\mathbf{w}^{j}_{q,l} \big\vert^{2}
\!\!\!+\! \sigma_m^2\!\! \Bigg). \label{eq:Rhat_rad_new}
\end{align}

To obtain a tractable surrogate, we linearize $\hat R^{\mathrm{rad}}_{m,q,l}[n]$ with respect to the sensing power variables around the current feasible point $\boldsymbol{\eta}^{s(f)}$, while treating the sensing beamformers $\{\mathbf{w}^{j}_{q,l}\}$ and the combiners $\{\mathbf{u}_{m,q,l}\}$ as fixed. Using the first-order Taylor expansion of the concave function $\log_2(\cdot)$ yields the global upper bound
\begin{align}
\hat R^{\mathrm{rad}}_{m,q,l}[n]\!\!
\le &
\hat R^{\mathrm{rad}(f)}_{m,q,l}[n]
+\!\!\!\!\!\sum_{j\in\mathcal M\setminus\{m\}}
\!\!\!\!B^{s(f)}_{j,m,q,l}[n]\,
\big(\eta^{s}_{j,q,l}[n]-\eta^{s(f)}_{j,q,l}[n]\big) \nonumber
\\ & \triangleq\hat R^{rp}_{m,q,l}[n], \label{eq:Rhat_rad_bound_new}
\end{align}
where $\hat R^{\mathrm{rad}(f)}_{m,q,l}[n]$ and $B^{s(f)}_{j,m,q,l}[n]$ are shown in~\eqref{eq:Rhat_rad_f_new} and ~\eqref{eq:B_rad_new}.
\begin{figure*}
\begin{gather}
\hat R^{\mathrm{rad}(f)}_{m,q,l}[n]
=\log_2\!\Bigg(
\sum_{j\in\mathcal M\setminus\{m\}}\sum_{k\in\mathcal K}
\eta^{s(f)}_{j,q,l}[n]\;
\big\vert\mathbf{u}_{m,q,l}^{H}\mathbf{A}_{j,m,k,q,l}\mathbf{w}^{j}_{q,l} \big\vert^{2}
+\sigma_m^2 \Bigg), \label{eq:Rhat_rad_f_new}
\\
B^{s(f)}_{j,m,q,l}[n]=\log_2(e)\;
\frac{\displaystyle\sum_{k\in\mathcal K}
\big\vert\mathbf{u}_{m,q,l}^{H}\mathbf{A}_{j,m,k,q,l}\mathbf{w}^{j}_{q,l} \big\vert^{2}}
{\displaystyle
\sum_{j'\in\mathcal M\setminus\{m\}}\sum_{k\in\mathcal K}
\eta^{s(f)}_{j',q,l}[n]\;
\big\vert\mathbf{u}_{m,q,l}^{H}\mathbf{A}_{j',m,k,q,l}\mathbf{w}^{j'}_{q,l} \big\vert^{2}
+\sigma_m^2}. \label{eq:B_rad_new}
\end{gather}
\end{figure*}

Replacing $\hat R^{\mathrm{rad}}_{m,q,l}[n]$ in \eqref{eq:R_rad_new}
with $\hat R^{rp}_{m,q,l}[n]$ gives the concave lower bound shown as follows
\begin{align}
&R^{\mathrm{rad,lb}}_{m,q,l}[n] \nonumber\\
= &\log_2 \Bigg(\!\!
\sum_{j\in\mathcal M}\sum_{k\in\mathcal K}
\eta^{s}_{j,q,l}[n]
\big\vert\mathbf{u}_{m,q,l}^{H}\mathbf{A}_{j,m,k,q,l}\mathbf{w}^{j}_{q,l} \big\vert^{2}
\!\!+\!\sigma_m^2\!\! \Bigg) \nonumber \\&- \hat R^{rp}_{m,q,l}[n],
\end{align}
which is used in the SCA subproblem for the sensing part.

Thus, constraint \eqref{eq:problem1_b} can be rewritten as the inequality in \eqref{eq:R_rad_geq}.
\begin{figure*}
    \begin{align}
        \sum_{m \in \mathcal{M}}\sum_{n \in \mathcal{N}} \sum^{Q}_{q=1} \sum^{L}_{l=1} \frac{\delta[n]}{QL} \cdot \left( \log_2 \left(\sum_{k \in \mathcal{K}}\sum_{j\in \mathcal{M}} \eta^{s}_{j,q,l}[n]
        \big\vert\mathbf{u}_{m,q,l}^{H}\mathbf{A}_{j,m,k,q,l}\mathbf{w}^{j}_{q,l} \big\vert^{2}
        + \sigma_{m}^2 \right) - \hat{R}^{rp}_{m,q,l}[n] \right) \geq R^{\mathrm{MI}}_{\min}.
    \label{eq:R_rad_geq}
    \end{align}
\end{figure*}
The surrogate constraint in \eqref{eq:R_rad_geq} is a convex constraint, namely a superlevel-set constraint of a concave function, and \eqref{eq:problem1_d} is also convex. Therefore, Problem $(\text{P}2.2)$ can be approximated at each SCA iteration by the following convex problem:
\begin{subequations}
\label{eq:problem2_3}
\begin{align}
    (\text{P}2.3): & \min_{\{\boldsymbol{\eta}^{s}[n]\}} \sum_{n \in \mathcal{N}}\sum_{m\in\mathcal M}\sum_{q=1}^{Q}\sum_{l=1}^{L}\eta^{s}_{m,q,l}[n] \nonumber\\
    \text{s.t.} \quad 
    & \eqref{eq:R_rad_geq},\eqref{eq:problem1_d}. \nonumber
\end{align}
\end{subequations}
where constraints \eqref{eq:R_rad_geq} and \eqref{eq:problem1_d} are convex, so the optimization problem is convex and can be solved using convex optimization tools. Note that Problem $(\text{P}2.3)$ optimizes an auxiliary sensing-power criterion for selecting a feasible sensing-power allocation with low power cost, and it does not directly improve the communication-rate objective of Problem $(\text{P}1)$.

\subsection{Sensing Duration Ratio Optimization}

Given the power allocation \( \boldsymbol{\eta}^{c} \), \( \boldsymbol{\eta}^{s} \), and the UAV trajectory \( \mathbf{q}_{k}[n] \), the optimization subproblem for the sensing duration ratio \( \delta[n] \) is formulated as follows
\begin{subequations}
    \label{eq:problem3}
    \begin{align}
    (\text{P}3): & \max_{\{\delta[n]\}} \sum_{n \in \mathcal{N}} R[n] \label{eq:problem3_a}\\
    \text{s.t.} \quad 
    & \eqref{eq:problem1_b}, \eqref{eq:problem1_d}, \eqref{eq:problem1_e}. \nonumber
    \end{align}
\end{subequations}

With $\boldsymbol{\eta}^{c}$, $\boldsymbol{\eta}^{s}$, and $\{\mathbf{q}_k[n]\}$ fixed, the objective $\sum_{n\in\mathcal N}R[n]$ is affine in $\delta[n]$ through the factor $1-\delta[n]$, and constraints \eqref{eq:problem1_b}, \eqref{eq:problem1_d}, and \eqref{eq:problem1_e} are also affine in $\delta[n]$. Therefore, Problem $(\text{P}3)$ is a linear program and can be solved efficiently by CVX.

\subsection{UAV Trajectory Optimization}

Given the power allocation $\boldsymbol{\eta}^{c}$, $\boldsymbol{\eta}^{s}$ and the sensing duration ratio $\delta[n]$, the UAV trajectory optimization can be formulated as follows
\begin{subequations}
    \label{eq:problem4}
    \begin{align}
    (\text{P}4): & \max_{\{\mathbf{q}_{k}[n]\}} \sum_{n \in \mathcal{N}} R[n] \label{eq:problem4_a}\\
    \text{s.t.} \quad 
    & \eqref{eq:problem1_b}, \eqref{eq:problem1_f}, \eqref{eq:problem1_g}, \eqref{eq:problem1_h},  \eqref{eq:problem1_i}. \nonumber
    \end{align}
\end{subequations}

The objective in \eqref{eq:problem4_a} and constraints \eqref{eq:problem1_b} and \eqref{eq:problem1_i} are non-convex with respect to $\{\mathbf q_k[n]\}$, and are handled by successive convex approximation. At outer iteration $f$, let $\{\hat{\mathbf q}_k^{(f-1)}[n]\}$ denote the current feasible trajectory and $\{\mathbf q_k[n]\}$ the updated variables. We first consider the collision-avoidance constraint \eqref{eq:problem1_i}. Since $\|\mathbf q_j[n]-\mathbf q_i[n]\|^2$ is convex, its first-order Taylor expansion at $\hat{\mathbf q}^{(f-1)}$ gives a global affine under-estimator. Replacing the original term with this under-estimator yields a conservative inner approximation that preserves safety at each SCA iteration. Hence, \eqref{eq:problem1_i} is convexified as
\begin{align}
    -\Vert \hat{\mathbf{q}}_j^{(f-1)}[n] - \hat{\mathbf{q}}_i^{(f-1)}[n] \Vert^2 
    + 2 \big(\hat{\mathbf{q}}_j^{(f-1)}[n] - \hat{\mathbf{q}}_i^{(f-1)}[n]\big)^T \nonumber \\ \cdot(\mathbf{q}_j[n] - \mathbf{q}_i[n]) 
    \geq D_{\min}^2 - (H_j - H_i)^2.
    \label{eq:q}
\end{align}

To make the trajectory dependence explicit, we separate the effective channel gain into a distance-related path-loss term and an array-response-related directional term. For fixed beamformers, $\left|\big(\mathbf h^c_{m,k}(\mathbf q_k[n],\mathbf v_m)\big)^H\mathbf w^c_{m,i}[n]\right|^2$ is the product of $\beta_0/(H_k^2+\|\mathbf q_k[n]-\mathbf v_m\|^2)$ and a directional alignment term. We therefore define
\begin{align}
    g_{m,k,i}(\mathbf{q}_{k}[n],\mathbf{v}_{m})
    = &\vert \mathbf{a}^{H}_{m,k}(\mathbf{q}_{k}[n],\mathbf{v}_{m}) \mathbf{w}^{c}_{m,i}[n] \vert^{2} \nonumber\\
    = &(\mathbf{w}^{c}_{m,i}[n])^{H}\mathbf{A}_{m,k}(\mathbf{q}_{k}[n],\mathbf{v}_{m}) \mathbf{w}^{c}_{m,i}[n],
\end{align}
where
\begin{align}
    \mathbf{A}_{m,k}(\mathbf{q}_{k}[n],\mathbf{v}_{m})
    = \mathbf{a}_{m,k}(\mathbf{q}_{k}[n],\mathbf{v}_{m})\mathbf{a}^{H}_{m,k}(\mathbf{q}_{k}[n],\mathbf{v}_{m}).
\end{align}
For fixed beamformers, $g_{m,k,i}(\mathbf q_k[n],\mathbf v_m)$ varies smoothly with $\mathbf q_k[n]$ through the geometry-dependent steering vector and is therefore suitable for first-order linearization.

To construct the affine surrogate \eqref{eq:g_affine_outer}, we compute the gradient of $g_{m,k,i}$ with respect to the UAV horizontal position. Since $g_{m,k,i}$ depends on $\mathbf q_k[n]$ through the array steering vector, the gradient is obtained by the chain rule from the horizontal coordinates to the direction cosines and then to the array response. For completeness and reproducibility, the intermediate derivatives of the direction cosines and the UPA steering vector are listed below.

For the explicit gradient expressions below, we specialize to the equal-altitude case $H_k=H$,$\forall k$, which is also the implementation adopted in the simulations. The same derivation extends to unequal but constant altitudes by retaining the corresponding $H_k$ term in $D_{m,k}$ and in the derivatives of the direction cosines. Define $\bar{\mathbf{q}}_{k}[n] = [q_{k,x}[n], q_{k,y}[n], H]$, $\bar{\mathbf{v}}_{m} = [v_{m,x}, v_{m,y}, 0]$, and $D_{m,k}(\bar{\mathbf{q}}_{k}[n], \bar{\mathbf{v}}_{m}) = \Vert \bar{\mathbf{q}}_{k}[n] - \bar{\mathbf{v}}_{m} \Vert$. The direction cosines are $\Phi(\bar{\mathbf{q}}_{k}[n], \bar{\mathbf{v}}_{m}) = (q_{k,x}[n] - v_{m,x})/D_{m,k}(\bar{\mathbf{q}}_{k}[n], \bar{\mathbf{v}}_{m})$ and $\Omega(\bar{\mathbf{q}}_{k}[n], \bar{\mathbf{v}}_{m}) = (q_{k,y}[n] - v_{m,y})/D_{m,k}(\bar{\mathbf{q}}_{k}[n], \bar{\mathbf{v}}_{m})$, whose partial derivatives are
\begin{align}
    \frac{\partial \Phi(\bar{\mathbf{q}}_{k}[n], \bar{\mathbf{v}}_{m})}{\partial q_{k,x}[n]}=&\frac{(q_{k,y}[n] - v_{m,y})^{2} + H^{2}}{D^{3}_{m,k}(\bar{\mathbf{q}}_{k}[n], \bar{\mathbf{v}}_{m})}, \\
    \frac{\partial \Omega(\bar{\mathbf{q}}_{k}[n], \bar{\mathbf{v}}_{m})}{\partial q_{k,y}[n]}=&\frac{(q_{k,x}[n] - v_{m,x})^{2} + H^{2}}{D^{3}_{m,k}(\bar{\mathbf{q}}_{k}[n], \bar{\mathbf{v}}_{m})},\\
    \frac{\partial \Phi(\bar{\mathbf{q}}_{k}[n], \bar{\mathbf{v}}_{m})}{\partial q_{k,y}[n]}=&-\frac{(q_{k,x}[n] - v_{m,x})(q_{k,y}[n] - v_{m,y})}{D^{3}_{m,k}(\bar{\mathbf{q}}_{k}[n], \bar{\mathbf{v}}_{m})}\nonumber\\
    =&\frac{\partial \Omega(\bar{\mathbf{q}}_{k}[n], \bar{\mathbf{v}}_{m})}{\partial q_{k,x}[n]}.
\end{align}

Denote the index vectors as $\mathbf{t}_{x}=[0,1,\ldots,N_{px}-1]$, $\mathbf{t}_{y}=[0,1,\ldots,N_{py}-1]$, then we have
\begin{align}
    \frac{\partial\mathbf{a}_{m,k}(\mathbf{q}_{k}[n],\mathbf{v}_{m})}{\partial \Phi(\bar{\mathbf{q}}_{k}[n], \bar{\mathbf{v}}_{m})}=& \left(-\jmath\frac{2\pi}{\lambda}d_{x} \text{diag}(\mathbf{t}_{x})\mathbf{a}_{m,k,x}(\mathbf{q}_{k}[n],\mathbf{v}_{m})\right) \nonumber \\ &\otimes \mathbf{a}_{m,k,y}(\mathbf{q}_{k}[n],\mathbf{v}_{m}), \\
    \frac{\partial\mathbf{a}_{m,k}(\mathbf{q}_{k}[n],\mathbf{v}_{m})}{\partial \Omega(\bar{\mathbf{q}}_{k}[n], \bar{\mathbf{v}}_{m})}=& \mathbf{a}_{m,k,x}(\mathbf{q}_{k}[n],\mathbf{v}_{m}) \nonumber \\ &\otimes \left(-\jmath\frac{2\pi}{\lambda}d_{y} \text{diag}(\mathbf{t}_{y})\mathbf{a}_{m,k,y}(\mathbf{q}_{k}[n],\mathbf{v}_{m})\right).
\end{align}

Using the Kronecker structure of the UPA, the derivatives with respect to the horizontal UAV coordinates are
\begin{align}
    \frac{\partial\mathbf{a}_{m,k}(\mathbf{q}_{k}[n],\mathbf{v}_{m})}{\partial q_{k,x}[n]} = &\left(\!\! \frac{\partial\mathbf{a}_{m,k,x}(\mathbf{q}_{k}[n],\mathbf{v}_{m})}{\partial \Phi(\bar{\mathbf{q}}_{k}[n], \bar{\mathbf{v}}_{m})} \! \cdot \! \frac{\partial \Phi(\bar{\mathbf{q}}_{k}[n], \bar{\mathbf{v}}_{m})}{\partial q_{k,x}[n]}\!\!\right) \nonumber \\ 
    &\otimes \mathbf{a}_{m,k,y}(\mathbf{q}_{k}[n],\mathbf{v}_{m}) \nonumber \\
    & + \mathbf{a}_{m,k,x}(\mathbf{q}_{k}[n],\mathbf{v}_{m}) \nonumber \\
    &\otimes \! \! \left( \!\!\frac{\partial\mathbf{a}_{m,k,y}(\mathbf{q}_{k}[n],\mathbf{v}_{m})}{\partial \Omega(\bar{\mathbf{q}}_{k}[n], \bar{\mathbf{v}}_{m})} \! \cdot \! \frac{\partial \Omega(\bar{\mathbf{q}}_{k}[n], \bar{\mathbf{v}}_{m})}{\partial q_{k,x}[n]}\!\!\right),
\end{align}
\begin{align}
    \frac{\partial\mathbf{a}_{m,k}(\mathbf{q}_{k}[n],\mathbf{v}_{m})}{\partial q_{k,y}[n]} = &\left( \!\!\frac{\partial\mathbf{a}_{m,k,x}(\mathbf{q}_{k}[n],\mathbf{v}_{m})}{\partial \Phi(\bar{\mathbf{q}}_{k}[n], \bar{\mathbf{v}}_{m})} \! \cdot \! \frac{\partial \Phi(\bar{\mathbf{q}}_{k}[n], \bar{\mathbf{v}}_{m})}{\partial q_{k,y}[n]}\!\!\right) \nonumber \\ 
    &\otimes \mathbf{a}_{m,k,y}(\mathbf{q}_{k}[n],\mathbf{v}_{m}) \nonumber \\
    & + \mathbf{a}_{m,k,x}(\mathbf{q}_{k}[n],\mathbf{v}_{m}) \nonumber \\ 
    & \otimes \! \! \left( \!\! \frac{\partial\mathbf{a}_{m,k,y}(\mathbf{q}_{k}[n],\mathbf{v}_{m})}{\partial \Omega(\bar{\mathbf{q}}_{k}[n], \bar{\mathbf{v}}_{m})} \! \cdot \! \frac{\partial \Omega(\bar{\mathbf{q}}_{k}[n], \bar{\mathbf{v}}_{m})}{\partial q_{k,y}[n]}\!\!\right).
\end{align}

Therefore, for $q_{k,d}[n] \in \{ q_{k,x}[n], q_{k,y}[n] \}$, we have
\begin{align}
    & \frac{\partial g_{m,k,i}(\mathbf{q}_{k}[n],\mathbf{v}_{m})}{\partial q_{k,d}[n]} \nonumber \\\!\! = & 2 \text{Re} \! \left\{ \! \! \left(\!\!\frac{\partial \mathbf{a}_{m,k}(\!\mathbf{q}_{k}[n],\!\!\mathbf{v}_{m}\!)}{\partial q_{k,d}[n]}\!\!\right)^{\!H} \!\!\!\! \mathbf{w}^{c}_{m,i}\![n]
    (\mathbf{w}^{c}_{m,i}\![n])^{H} \mathbf{a}_{m,k}(\!\mathbf{q}_{k}[n],\!\!\mathbf{v}_{m}\!)\!\!\right\}.
\end{align}

Substituting $\partial\mathbf{a}_{m,k}(\mathbf{q}_{k}[n],\mathbf{v}_{m})/\partial q_{k,x}[n]$ yields the explicit gradient $\nabla_{\mathbf{q}}g_{m,k,i}(\mathbf{q}_{k}[n],\mathbf{v}_{m})$.

Given this gradient, at the $f$-th outer SCA iteration we linearize $g_{m,k,i}(\mathbf{q}_{k}[n],\mathbf{v}_{m})$ at the current reference point $\hat{\mathbf{q}}^{(f-1)}_{k}[n]$ as
\begin{align}
    &g_{m,k,i}(\mathbf{q}_{k}[n],\mathbf{v}_{m}) \approx \tilde{g}^{(f)}_{m,k,i}(\mathbf{q}_{k}[n],\mathbf{v}_{m}) \nonumber \\
    = &g_{m,k,i}(\hat{\mathbf{q}}^{(f-1)}_{k}[n],\mathbf{v}_{m}) \nonumber \\ 
    &+ \nabla_{\mathbf{q}}g^{T}_{m,k,i}(\hat{\mathbf{q}}^{(f-1)}_{k}[n],\mathbf{v}_{m})\left( \mathbf{q}_{k}[n] - \hat{\mathbf{q}}^{(f-1)}_{k}[n] \right),
    \label{eq:g_affine_outer}
\end{align}
which is an affine approximation in $\mathbf{q}_{k}[n]$ used inside the SCA update.

To keep \eqref{eq:g_affine_outer} accurate and ensure monotonic improvement, we impose the trust-region constraint around the current outer-iteration trajectory:
\begin{align}
    \|\mathbf{q}_{k}[n]-\hat{\mathbf{q}}_{k}^{(f-1)}[n]\| \le \psi^{(f)}, \quad \forall k\in\mathcal{K},\ n\in\mathcal{N},
    \label{eq:trust_region}
\end{align}
where $\psi^{(f)}$ is the nominal trust-region radius at outer iteration $f$. In the inner alternating updates, a local radius initialized from $\psi^{(f)}$ may be further shrunk for acceptance control, while the trust-region center remains fixed at $\hat{\mathbf{q}}^{(f-1)}$.

Meanwhile, the path-loss term contains the reciprocal factor $1/(H_k^2+\|\mathbf q_k[n]-\mathbf v_m\|^2)$, which is coupled with the trajectory inside the rate logarithms. To decouple it from the quadratic distance term, we introduce an epigraph variable $z_{m,k}[n]$ that upper-bounds the squared slant range. Since the reciprocal function is decreasing on the positive real line, the constraint $z_{m,k}[n]\ge H_k^2+\|\mathbf q_k[n]-\mathbf v_m\|^2$ yields the conservative lower bound $1/(H_k^2+\|\mathbf q_k[n]-\mathbf v_m\|^2)\ge 1/z_{m,k}[n]$, which is suitable for constructing concave lower-bound surrogates. We further approximate $1/z_{m,k}[n]$ by its first-order Taylor lower bound at the previous inner iterate $z^{(r-1)}_{m,k}[n]$:
\begin{align}
\ell^{(r-1)}_{m,k}(z_{m,k}[n])
\triangleq
\frac{1}{z^{(r-1)}_{m,k}[n]}
-
\frac{z_{m,k}[n]-z^{(r-1)}_{m,k}[n]}{\big(z^{(r-1)}_{m,k}[n]\big)^2},
\label{eq:ell_lower}
\end{align}
which is tight and gradient-matching at $z_{m,k}[n]=z^{(r-1)}_{m,k}[n]$.

To avoid bilinear coupling between $\mathbf q_k[n]$ and $z_{m,k}[n]$ in the surrogate SINR expressions, we adopt a two-block update: the $q$-step fixes $z$ and treats the reciprocal term as constant, while the $z$-step fixes $\mathbf q$ and updates the affine reciprocal surrogate. The resulting ``signal+interference'' and ``interference-only'' sums for UAV $k$ in slot $n$ are
\begin{align}
  S_k^{(q)}[n] &\triangleq
    \sum_{m\in\mathcal M}\sum_{i\in\mathcal K}
      \beta_0\,\eta^c_{m,i}[n]\,
      \frac{\tilde g^{(f)}_{m,k,i}(\mathbf q_k[n])}{z^{(r-1)}_{m,k}[n]},
      \label{eq:Sk_qstep}\\
  I_k^{(q)}[n] &\triangleq
    \sum_{m\in\mathcal M}\sum_{i\in\mathcal K\setminus\{k\}}
      \beta_0\,\eta^c_{m,i}[n]\,
      \frac{\tilde g^{(f)}_{m,k,i}(\mathbf q_k[n])}{z^{(r-1)}_{m,k}[n]},
      \label{eq:Ik_qstep}\\[2mm]
  S_k^{(z)}[n] &\triangleq
    \sum_{m\in\mathcal M}\sum_{i\in\mathcal K}
      \beta_0\,\eta^c_{m,i}[n]\,
      \tilde g^{(f)}_{m,k,i}\!\big(\mathbf q_k^{(r)}[n]\big)\,
      \ell^{(r-1)}_{m,k}(z_{m,k}[n]),
      \label{eq:Sk_zstep}\\
  I_k^{(z)}[n] &\triangleq
    \sum_{m\in\mathcal M}\sum_{i\in\mathcal K\setminus\{k\}} \!\!\!\!
      \beta_0\,\eta^c_{m,i}[n]\,
      \tilde g^{(f)}_{m,k,i}\!\big(\mathbf q_k^{(r)}[n]\big)\,
      \ell^{(r-1)}_{m,k}(z_{m,k}[n]).
      \label{eq:Ik_zstep}
\end{align}

For the rate function $R_k[n]=\log_2(S_k[n]+\sigma_k^2)-\log_2(I_k[n]+\sigma_k^2)$, we keep the first term and upper-bound the second term by its tangent, using the concavity of the logarithm. The approximation is taken at the reference points $y_k^{(q)}[n] = I_k^{(q,\text{ref})}[n] + \sigma_k^2$ and $y_k^{(z)}[n] = I_k^{(z,\text{ref})}[n] + \sigma_k^2$, where $I_k^{(q,\text{ref})}[n]$ and $I_k^{(z,\text{ref})}[n]$ are evaluated at the previous inner iterate so that the bound is tight at the current operating point. The resulting per-UAV concave surrogates for the two inner steps are given by
\begin{align}
  \widetilde R_k^{(q)}[n]
  =& \log_2\!\big(S_k^{(q)}[n]+\sigma_k^2\big) \nonumber \\
   &-\Big[\log_2 y_k^{(q)}[n]
     + \tfrac{I_k^{(q)}[n]+\sigma_k^2-y_k^{(q)}[n]}{y_k^{(q)}[n]\ln 2}\Big], \label{eq:Rk_q}\\
  \widetilde R_k^{(z)}[n]
  =& \log_2\!\big(S_k^{(z)}[n]+\sigma_k^2\big) \nonumber \\
   &-\Big[\log_2 y_k^{(z)}[n]
     + \tfrac{I_k^{(z)}[n]+\sigma_k^2-y_k^{(z)}[n]}{y_k^{(z)}[n]\ln 2}\Big]. \label{eq:Rk_z}
\end{align}

Both $\widetilde R_k^{(q)}[n]$ and $\widetilde R_k^{(z)}[n]$ are concave in their respective decision variables, since
$S_k^{(\cdot)}$ is affine and the bracketed term is an affine upper bound of $\log(\cdot)$.  
The slot-weighted sum surrogate is
\begin{align}
  \widetilde R^{(q)}[n] \triangleq \sum_{k\in\mathcal K} (1-\delta[n])\,\widetilde R_k^{(q)}[n],\\
  \widetilde R^{(z)}[n] \triangleq \sum_{k\in\mathcal K} (1-\delta[n])\,\widetilde R_k^{(z)}[n].
\end{align}

With these concave surrogates established, we are now positioned to solve the trajectory optimization Problem (P4). To keep the actual block updates consistent with the MI-constrained formulation in Problem (P4), the radar-MI constraint \eqref{eq:problem1_b} is enforced in both the $q$-step and the $z$-step through block-wise concave lower bounds constructed at the current iterate. Specifically, we use
\begin{align}
\underline{R}^{\mathrm{rad},(q)}\!\left(\{\mathbf q_k[n]\};\{\mathbf q_k^{(r-1)}[n]\},\{z_{m,k}^{(r-1)}[n]\}\right)\ge R_{\min}^{\mathrm{MI}},
\label{eq:Rrad_traj_q_lb}
\end{align}
for the $q$-step, and
\begin{align}
\underline{R}^{\mathrm{rad},(z)}\!\left(\{z_{m,k}[n]\};\{\mathbf q_k^{(r)}[n]\},\{z_{m,k}^{(r-1)}[n]\}\right)\ge R_{\min}^{\mathrm{MI}},
\label{eq:Rrad_traj_z_lb}
\end{align}
for the $z$-step, where $\underline{R}^{\mathrm{rad},(q)}(\cdot)$ and $\underline{R}^{\mathrm{rad},(z)}(\cdot)$ denote the SCA-based concave lower-bound surrogates of the radar MI with respect to the current block variable. These lower bounds are obtained by applying the same SCA principle used in the sensing-power subproblem: we keep the concave log term and linearize the concave interference-related log term at the current iterate, which yields a concave lower bound with respect to the active block variable. We then employ an alternating optimization strategy that decomposes the problem into two convex subproblems: one for updating the UAV trajectories $\{\mathbf q_k[n]\}$ (the $q$-step), and another for updating the epigraph variables $\{z_{m,k}[n]\}$ (the $z$-step).

\subsubsection*{$q$-step}
With $z=z^{(r-1)}$ fixed, the trajectory subproblem is
\begin{subequations}
    \label{eq:problem4_1}
    \begin{align}
    (\text{P}4.1): & \max_{\{\mathbf q_k[n]\}} \sum_{n\in\mathcal N} \widetilde R^{(q)}[n] \\
    \text{s.t.}\quad 
    & H_k^2 + \|\mathbf q_k[n]-\mathbf v_m\|^2 \le z^{(r-1)}_{m,k}[n],~~\forall m,k,n, \\
    & \|\mathbf q_k[n]-\hat{\mathbf q}^{(f-1)}_k[n]\| \le \psi_{\mathrm{loc}}^{(r)}, \quad \forall k\in\mathcal K,\ n\in\mathcal N, \nonumber \\
    & \eqref{eq:problem1_f}, \eqref{eq:problem1_g}, \eqref{eq:problem1_h}, \eqref{eq:q}, \eqref{eq:Rrad_traj_q_lb}, \nonumber
    \end{align}
\end{subequations}
where $S_k^{(q)}[n]$ and $I_k^{(q)}[n]$ follow \eqref{eq:Sk_qstep}-\eqref{eq:Ik_qstep}. The objective is concave in $\{\mathbf q_k[n]\}$, and all constraints are convex, including the radar-MI surrogate constraint \eqref{eq:Rrad_traj_q_lb} (a superlevel-set constraint of a concave lower bound). Hence, Problem $(\text{P}4.1)$ is convex.

\subsubsection*{$z$-step}
With $\mathbf q=\mathbf q^{(r)}$ fixed, the epigraph update is
\begin{subequations}
    \label{eq:problem4_2}
    \begin{align}
    (\text{P}4.2): & \max_{\{z_{m,k}[n]\}} \sum_{n\in\mathcal N} \widetilde R^{(z)}[n] \\
    \text{s.t.}\quad 
    & H_k^2 + \|\mathbf q^{(r)}_k[n]-\mathbf v_m\|^2 \le z_{m,k}[n],~~\forall m,k,n, \\
    & \eqref{eq:Rrad_traj_z_lb}, \nonumber
    \end{align}
\end{subequations}
where $S_k^{(z)}[n]$ and $I_k^{(z)}[n]$ use \eqref{eq:Sk_zstep}-\eqref{eq:Ik_zstep} together with the affine lower bound \eqref{eq:ell_lower}. The radar-MI constraint is enforced via the convex surrogate \eqref{eq:Rrad_traj_z_lb}, so Problem $(\text{P}4.2)$ is also convex.

\begin{algorithm}[t]
\caption{SCA-Based Trajectory Optimization with Alternating $q$/$z$ Updates}
\label{alg:corrected_sca}
\begin{algorithmic}[1]
\State Initialize UAV trajectory $\{\hat{\mathbf q}^{(0)}[n]\}$, outer trust-region radius $\psi^{(1)}$, and outer index $f=1$.
\Repeat
    \State Construct affine gain surrogate $\tilde g^{(f)}_{m,k,i}(\mathbf q_k[n])$ at $\{\hat{\mathbf q}^{(f-1)}[n]\}$.
    \State Initialize inner loop: $r=1$, $\mathbf q^{(0)}=\hat{\mathbf q}^{(f-1)}$, $z^{(0)}_{m,k}[n]=H_k^2+\|\mathbf q_k^{(0)}[n]-\mathbf v_m\|^2$, and $\psi_{\mathrm{loc}}^{(1)}=\psi^{(f)}$.
    \Repeat
        \State \textbf{$q$-step:} Solve the convex trajectory subproblem with fixed $z^{(r-1)}$ under the local trust region and the radar-MI surrogate constraint, and obtain $\mathbf q_{\mathrm{cand}}$.
        \If{$\sum_n R[n]\vert_{\mathbf q=\mathbf q_{\mathrm{cand}},\,z=z^{(r-1)}} \ge
        \sum_n R[n]\vert_{\mathbf q=\mathbf q^{(r-1)},\,z=z^{(r-1)}}$}
            \State $\mathbf q^{(r)} \gets \mathbf q_{\mathrm{cand}}$, \quad
            $\psi_{\mathrm{loc}}^{(r+1)} \gets \psi_{\mathrm{loc}}^{(r)}$.
        \Else
            \State $\mathbf q^{(r)} \gets \mathbf q^{(r-1)}$, \quad
            $\psi_{\mathrm{loc}}^{(r+1)} \gets \psi_{\mathrm{loc}}^{(r)}/2$.
        \EndIf
        \State \textbf{$z$-step:} Solve the convex $z$-subproblem with fixed $\mathbf q^{(r)}$ using \eqref{eq:ell_lower} and the radar-MI surrogate constraint to obtain $z^{(r)}$.
        \State $r \gets r+1$.
    \Until{inner improvement $<\varepsilon_{\mathrm{in}}$ or $\psi_{\mathrm{loc}}^{(r)}<\hat\varepsilon$}
    \State $\hat{\mathbf q}^{(f)} \gets \mathbf q^{(r-1)}$, \quad
    $\psi^{(f+1)} \gets \psi_{\mathrm{loc}}^{(r)}$, \quad
    $f \gets f+1$.
\Until{objective improvement $<\bar\varepsilon$}
\end{algorithmic}
\end{algorithm}

The proposed alternating SCA-based scheme transforms the original non-convex trajectory optimization into a sequence of tractable convex programs. By using concave surrogate objectives together with trust-region control and an acceptance rule, the method produces a non-decreasing accepted objective sequence while preserving feasibility of the convexified constraints. This framework therefore provides a practical and efficient approach for UAV trajectory design in ISAC networks, balancing communication throughput and sensing reliability.

\subsection{Computational Complexity, Scalability, and Convergence Discussion}
\label{subsec:complexity_convergence}

This subsection clarifies (i) how the proposed AO--SCA algorithm scales with the network size (numbers of UAVs, BSs, and sensing beams), (ii) its computational complexity per iteration, and (iii) its convergence behavior. These clarifications address the practicality and scalability concerns in larger or denser cooperative ISAC networks.

\subsubsection{Complexity bookkeeping and problem dimensions}
In each AO iteration, four variable blocks are updated: communication power, sensing power, time-division ratio, and UAV trajectory. Their dimensions scale as \(MKN\), \(MQLN\), \(N\), and \(2KN\), respectively, plus \(MKN\) epigraph variables in the trajectory module. Hence, the overall complexity grows polynomially with \(K\), \(M\), \(QL\), and \(N\), with the sensing-power block dominating for dense beam grids and the trajectory block dominating for large UAV fleets.

\subsubsection{Per-iteration complexity and dominant scaling}
After applying the proposed SCA surrogates, all block updates become convex programs and can be solved in polynomial time by standard interior-point methods. To characterize the dependence on \((K,M,Q,L,N)\), we use the standard complexity proxy \(\mathrm{poly}(n,m)\) for a convex program with \(n\) variables and \(m\) constraints.

Let \(I_c\) and \(I_s\) denote the SCA iteration numbers for the communication-power and sensing-power updates, respectively, within each AO iteration. Let \(I_{\mathrm{tr}}\) denote the number of outer SCA refinements in the trajectory module, indexed by \(f\) in Algorithm~\ref{alg:corrected_sca}, and let \(I_{\mathrm{in}}\) denote the number of inner alternating \(q\)/\(z\) updates, indexed by \(r\). Then the per-AO-iteration computational cost can be decomposed as follows:
\begin{align}
\mathcal{C}_{\mathrm{AO}}
= &\;
I_c \cdot \mathrm{poly}(n_c, \bar m_c)
+ I_s \cdot \mathrm{poly}(n_s, \bar m_s)
+ \mathrm{poly}(n_\delta, \bar m_\delta) \nonumber\\
&\;
+ I_{\mathrm{tr}} I_{\mathrm{in}}
\Big(
\mathrm{poly}(n_q, \bar m_q) + \mathrm{poly}(n_z, \bar m_z)
\Big),
\label{eq:complexity_AO}
\end{align}
where $\bar m_c,\bar m_s,\bar m_\delta,\bar m_q,\bar m_z$ denote the numbers of \emph{structural} convex constraints excluding nonnegativity constraints, with the dominant scaling
\begin{align}
\bar m_c &= O(MN), \\
\bar m_s &= O(MN + 1), \\
\bar m_\delta &= O(MN + 1), \\
\bar m_q &= O(K^2 N + KN + MKN + 1), \\
\bar m_z &= O(MKN + 1).
\end{align}
If nonnegativity constraints are also counted, then $m_c=\bar m_c+n_c$, $m_s=\bar m_s+n_s$, and similarly for the other blocks. This bookkeeping highlights that the cumulative MI requirement contributes $O(1)$ global constraint, whereas the dominant growth with respect to $K$, $M$, $QL$, and $N$ mainly comes from the per-slot resource constraints and the multi-UAV safety constraints.

From \eqref{eq:complexity_AO}, the following scalability trends are immediate.

\textbf{Scaling with the number of UAVs $K$:}
The communication-power and trajectory subproblems scale with $K$ through $n_c = MKN$ and $n_q = 2KN$, and the number of safety constraints scales on the order of $O(K^2 N)$. Therefore, denser multi-UAV deployments increase the cost mainly via the trajectory module and the interference-coupled communication-power update.

\textbf{Scaling with the number of BSs $M$:}
All blocks that involve BS resources scale with $M$ through $n_c = MKN$, $n_s = MQLN$, and $n_z = MKN$. Increasing $M$ improves spatial diversity but also increases the coupling dimension in cooperative power and sensing allocation.

\textbf{Scaling with the number of sensing beams $QL$:}
The sensing-power update scales linearly in the number of beam directions through $n_s = MQLN$.
Thus, finer angular grids (larger $QL$) increase the sensing optimization dimension and lead to higher computational burden, reflecting a natural performance--complexity tradeoff between sensing resolution and optimization cost.

\textbf{Scaling with the mission length $N$:}
All blocks scale with $N$ because resource allocation and trajectory variables are defined per slot. In particular, $n_c$, $n_s$, $n_q$, and $n_z$ are all linear in $N$.
Therefore, for longer missions (larger $N$), a common practical approach is to use a receding-horizon implementation with a sliding window or to adopt a coarser slot discretization.

Moreover, compared with exhaustive search over the sensing/communication split and beam-sweeping patterns, the proposed AO--SCA method avoids combinatorial complexity by solving a sequence of convex programs. Hence, for fixed iteration budgets, the computational burden scales polynomially with the problem dimensions, making the approach suitable for mission-level offline planning or periodic re-optimization.

\subsubsection{Practical scalability and extensibility to larger networks}
The proposed AO--SCA framework is extensible to larger or denser cooperative ISAC networks because all subproblems remain convex after the same surrogate construction. Its computational cost still grows with $(K,M,QL,N)$, as discussed above. In practice, the burden can be reduced without changing the core formulation by: 
(i) parallelizing SINR/MI coefficient and gradient evaluation across BSs or time slots; 
(ii) warm-starting each convex subproblem from the previous AO iterate; 
(iii) limiting the numbers of AO and inner SCA iterations once the objective saturates; and 
(iv) adopting hierarchical or distributed implementations, such as slower trajectory updates, faster power/time-division updates, or clustered BS cooperation to reduce the effective $M$ in dense deployments. 
These strategies preserve the mathematical structure of the proposed AO--SCA method and are standard in large-scale cooperative optimization.

\subsubsection{Convergence discussion}
We briefly discuss the convergence of the proposed AO--SCA procedure.

\textbf{1) Monotonic improvement:}
In each AO iteration, the communication-power update maximizes a tight concave lower bound of the original sum-rate objective and therefore does not decrease the achieved sum rate. The sensing-power update solves an auxiliary feasibility-preserving minimization, and it does not decrease the communication objective because $\sum_{n\in\mathcal N}R[n]$ is independent of $\{\boldsymbol{\eta}^{s}[n]\}$ under the adopted TDD model. The time-division update is solved optimally as a linear program under fixed other blocks and is also non-decreasing. For the trajectory update, Algorithm~\ref{alg:corrected_sca} uses concave surrogate objectives together with a trust-region acceptance rule, so each accepted update preserves feasibility and improves or maintains the objective. Hence, the AO objective sequence is non-decreasing.

\textbf{2) Boundedness:}
The objective is upper bounded because the transmit powers satisfy \eqref{eq:problem1_c}--\eqref{eq:problem1_d} and $0 \le \delta[n] \le 1$. Therefore, the non-decreasing objective sequence converges to a finite limit.

\textbf{3) Convergence of the adopted AO--SCA procedure:}
The communication-power and sensing-power blocks use standard tight SCA surrogates at the current iterate, and the time-division block is solved exactly. For the trajectory block, Algorithm~\ref{alg:corrected_sca} adopts trust-region sequential convexification with an acceptance rule, so each accepted update is non-decreasing and remains feasible with respect to the convexified constraints. Therefore, together with the monotonicity and boundedness established above, the overall AO objective converges to a finite value. A full stationary-point proof for the trajectory block would require additional assumptions on the local surrogate quality within the trust region and is left for future work.

The proposed method is intended for mission-level planning and slot-level scheduling, and is therefore more suitable for offline optimization or periodic re-optimization, such as receding-horizon updates, than for symbol-level real-time control.

\section{Numerical Results}

\subsection{Simulation Setup}
In this section, we provide numerical results to demonstrate the effectiveness of our proposed joint optimization algorithm. We consider a scenario with $K=3$ UAVs and $M=3$ BSs. The total flight duration is $T=40$\,s, which is divided into $N=40$ time slots. Each BS is equipped with a UPA with $N_{px} \times N_{py} = 8 \times 8$ antenna elements, and the antenna spacing is set to half a wavelength, i.e., $d_x = d_y = \lambda/2$, with $\lambda = 0.1$\,m. The fixed altitude for all UAVs is set to $H_k = 80$\,m for $k \in \{1, 2, 3\}$.

To evaluate the performance of the proposed algorithm under different geometrical layouts, we consider two distinct position settings for the UAVs, while the locations of the BSs remain fixed. The locations of the three BSs are set at $\mathbf v_{1} = [150, 250]$\,m, $\mathbf{v}_{2} = [350, 100]$\,m, and $\mathbf{v}_{3} = [200, -75]$\,m. The other parameters remain consistent across both settings: the maximum flight speed of each UAV is $V_{\max} = 25$\,m/s, and the minimum required separation distance is $D_{\min}=20$\,m.

\textbf{Position Setting 1}: The initial positions of the three UAVs are $\mathbf{q}^{\mathrm{I}}_{1} = [0, 0]$\,m, $\mathbf{q}^{\mathrm{I}}_{2} = [0, 150]$\,m, and $\mathbf{q}^{\mathrm{I}}_{3} = [0, -150]$\,m. Their corresponding final positions are $\mathbf{q}^{\mathrm{F}}_{1} = [500, 0]$\,m, $\mathbf{q}^{\mathrm{F}}_{2} = [500, 200]$\,m, and $\mathbf{q}^{\mathrm{F}}_{3} = [500, -200]$\,m.

\textbf{Position Setting 2}: The initial positions are set to $\mathbf{q}^{\mathrm{I}}_{1} = [0, 250]$\,m, $\mathbf{q}^{\mathrm{I}}_{2} = [0, 100]$\,m, and $\mathbf{q}^{\mathrm{I}}_{3} = [0, -50]$\,m. The corresponding final positions are $\mathbf{q}^{\mathrm{F}}_{1} = [500, 250]$\,m, $\mathbf{q}^{\mathrm{F}}_{2} = [500, 100]$\,m, and $\mathbf{q}^{\mathrm{F}}_{3} = [500, -50]$\,m.

For the communication and sensing parameters, we normalize the channel gain at a reference distance by setting $\beta_0 = 1$ to focus on the relative performance of our algorithm. The maximum transmit powers for communication and sensing at each BS are $P^{c}_{\max} = 10$\,W and $P^{s}_{\max} = 10$\,W, respectively. The noise powers at the UAVs and BSs are set to $\sigma_k^2 = -30$ dBm and $\sigma_m^2 = -30$ dBm, respectively. The minimum total sensing MI requirement for the entire mission is set to $R_{\min}^{\mathrm{MI}} = 100$. The sensing grid is defined by $Q=7$ zenith angles and $L=16$ azimuth angles. Unless specified otherwise, these parameters are used in the following simulations. 

\subsection{Benchmark Schemes}
For a comprehensive performance evaluation, we compare our proposed joint optimization algorithm with four benchmark schemes designed to isolate the contributions of different optimization components.

\textbf{Benchmark 1: Static Trajectory with Optimized Resources}: 
In this baseline, the UAV trajectories are fixed and not optimized. Each UAV flies along a straight-line path from its initial position to its final position with a constant speed, i.e., the trajectory is fully determined by \( \mathbf{q}^{\mathrm{I}}_{k} \), \( \mathbf{q}^{\mathrm{F}}_{k} \), and the mission duration \(T\). Under these predetermined flight paths, we optimize only the resource allocation variables, including the communication power \(\{\boldsymbol{\eta}^{c}[n]\}\), sensing power \(\{\boldsymbol{\eta}^{s}[n]\}\), and time-division ratio \(\{\delta[n]\}\), to maximize the total communication sum rate subject to the sensing MI requirement. This benchmark isolates the gain brought purely by adaptive resource allocation when trajectory design is unavailable.

\textbf{Benchmark 2: Uniform Power Allocation with Optimized Trajectory and Time}: 
This benchmark evaluates the importance of adaptive power control. In this scheme, the communication and sensing powers are not optimized across users or beam directions. Specifically, the communication power at each BS is evenly distributed across UAVs, namely \( \eta^{c}_{m,k}[n] = P_{\max}^{c}/K \). For sensing, we impose uniform power across all beam directions within each slot, namely $\eta^{s}_{m,q,l}[n] = \bar{\eta}^{s}_{m}[n]$ for all $q, l$, where the common value $\bar{\eta}^{s}_{m}[n]$ may vary across slots and is chosen as the largest feasible uniform value under \eqref{eq:problem1_d}. Under this fixed uniform power-allocation structure, we optimize the UAV trajectories \(\{\mathbf{q}_{k}[n]\}\) together with the slot-dependent time-division ratio \(\{\delta[n]\}\). This benchmark highlights the performance gain achieved by adaptive communication and sensing power allocation.

\textbf{Benchmark 3: Uniform Time Division with Optimized Trajectory and Power}: 
In this benchmark, the sensing-communication time split is constrained to be identical across all time slots, namely \( \delta[n] = \delta_{\mathrm{opt}} \) for all \( n \in \mathcal{N} \), where the common ratio \( \delta_{\mathrm{opt}} \) is itself optimized. Meanwhile, the UAV trajectories \(\{\mathbf{q}_{k}[n]\}\), communication power allocation \(\{\boldsymbol{\eta}^{c}[n]\}\), and sensing power allocation \(\{\boldsymbol{\eta}^{s}[n]\}\) are jointly optimized. Compared with the proposed design, this benchmark removes the slot-by-slot flexibility of time scheduling and is used to demonstrate the benefit of dynamically adapting the sensing duration to time-varying geometry and channel conditions.

\textbf{Benchmark 4: Dynamic Best-Server Selection with Optimized Resources and Trajectory}: 
This benchmark is introduced as a strong literature-inspired non-cooperative baseline for multi-cell UAV networking scenarios \cite{9858656, cheng2024networked}. In this scheme, each UAV is associated with only one serving BS in each time slot according to the strongest instantaneous BS--UAV link under the current geometry. Hence, unlike the proposed cooperative multi-BS transmission design, no UAV can be jointly served by multiple BSs in the same slot. Under this single-serving-BS constraint, we still optimize the UAV trajectories \(\{\mathbf{q}_{k}[n]\}\), communication power allocation, sensing power allocation, and time-division ratio \(\{\delta[n]\}\) to maximize the total communication sum rate while satisfying the same sensing MI constraint. Therefore, this benchmark preserves the main optimization structure while removing the gain brought by cooperative multi-BS serving, enabling a fair evaluation of the benefit of BS cooperation.

\subsection{Trajectory and Slot-Level Behavior Analysis}

\begin{figure*}[t]
    \centering
    \begin{subfigure}[b]{0.245\textwidth}
        \centering
        \includegraphics[width=\textwidth]{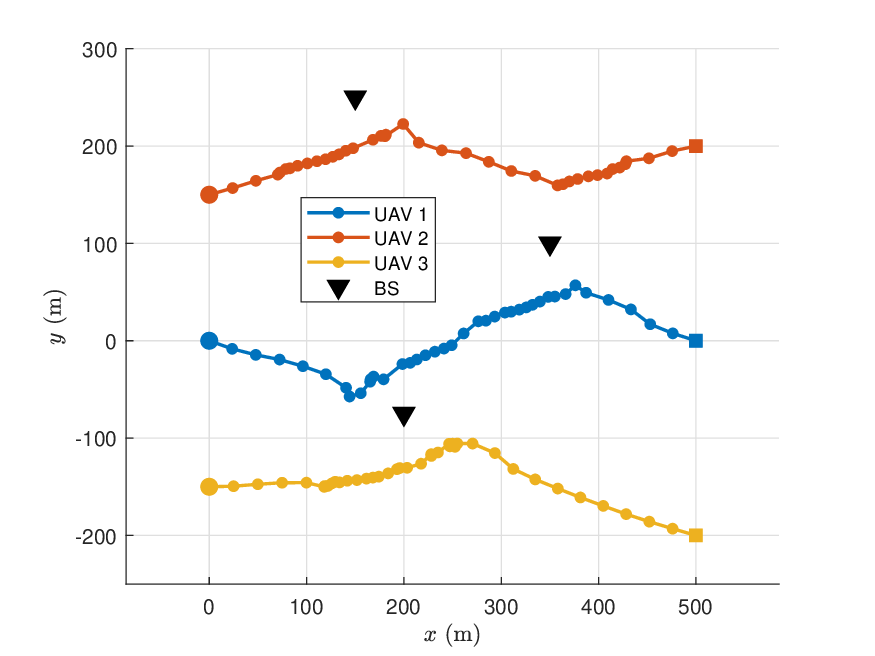}
        \caption{Setting 1, $R_{\min}^{\mathrm{MI}}=60$\,bits}
        \label{fig:trajectory_image1}
    \end{subfigure}
    \hfill
    \begin{subfigure}[b]{0.245\textwidth}
        \centering
        \includegraphics[width=\textwidth]{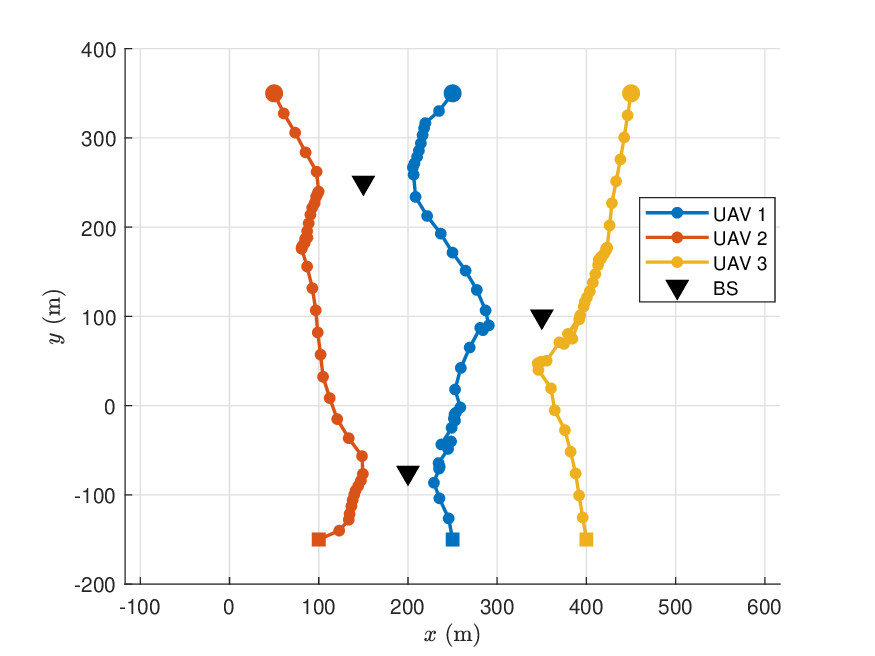}
        \caption{Setting 2, $R_{\min}^{\mathrm{MI}}=60$\,bits}
        \label{fig:trajectory_image2}
    \end{subfigure}
    \hfill
    \begin{subfigure}[b]{0.245\textwidth}
        \centering
        \includegraphics[width=\textwidth]{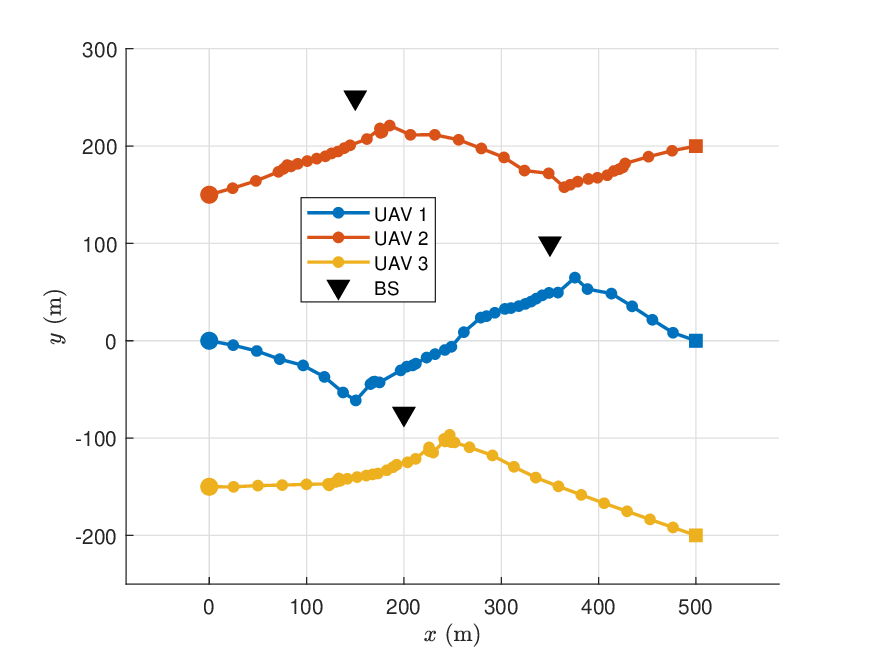}
        \caption{Setting 1, $R_{\min}^{\mathrm{MI}}=100$\,bits}
        \label{fig:trajectory_image3}
    \end{subfigure}
    \hfill
    \begin{subfigure}[b]{0.245\textwidth}
        \centering
        \includegraphics[width=\textwidth]{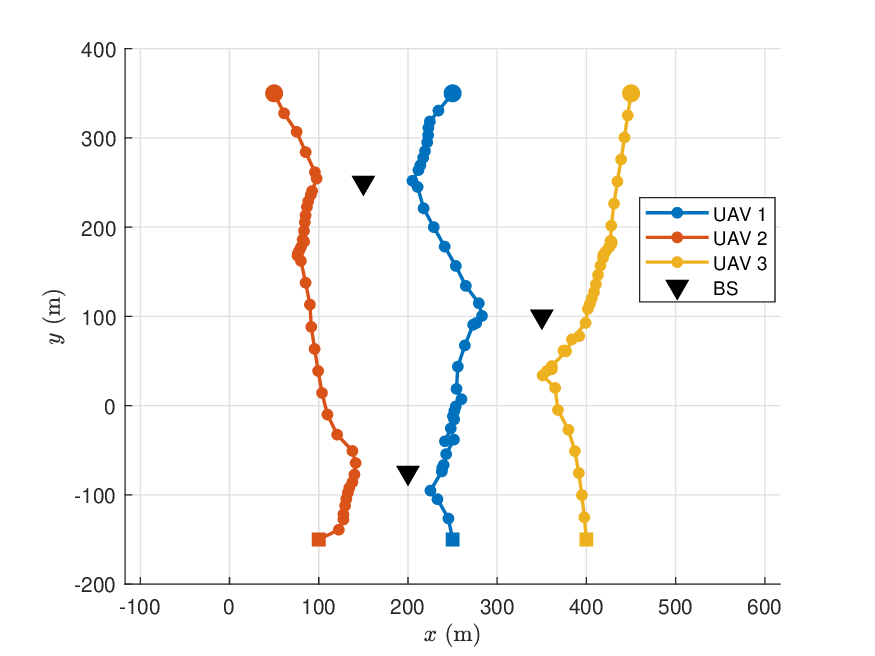}
        \caption{Setting 2, $R_{\min}^{\mathrm{MI}}=100$\,bits}
        \label{fig:trajectory_image4}
    \end{subfigure}

    \caption{Optimized UAV trajectories under different position settings and MI thresholds.}
    \label{fig:optimized_trajectory}
\end{figure*}

Fig.~\ref{fig:optimized_trajectory} shows the optimized UAV trajectories for both position settings under varying minimum sensing MI constraints. Across all subfigures, the UAVs adjust speed dynamically. Trajectory markers, which indicate the position at each time slot, are sparse when the UAVs are far from any BS, revealing higher speeds in weak-channel regions; they become dense near the BSs, indicating slower motion. This behavior is effective since dwelling in strong-channel areas yields higher data rates and advances the goal of maximizing the sum communication rate. The MI requirement also shapes the paths. Fig.~\ref{fig:trajectory_image1} and Fig.~\ref{fig:trajectory_image3} exhibit similar overall geometry, yet the stricter target \(R_{\min}^{\mathrm{MI}}=100\) bits moves the trajectories slightly closer to the BSs than \(R_{\min}^{\mathrm{MI}}=60\) bits, which strengthens the received echoes and improves sensing quality to meet the higher target. The UAVs maintain a clear standoff distance rather than flying arbitrarily close to the BSs. Although shorter range strengthens the desired signal, it raises inter-UAV interference, degrades the SINR of other links, and can reduce the sum communication rate. The optimized trajectories therefore balance signal strength and interference containment to enhance both sensing reliability and communication throughput.

\begin{figure}[h]
    \centering
    \includegraphics[width=0.4\textwidth]{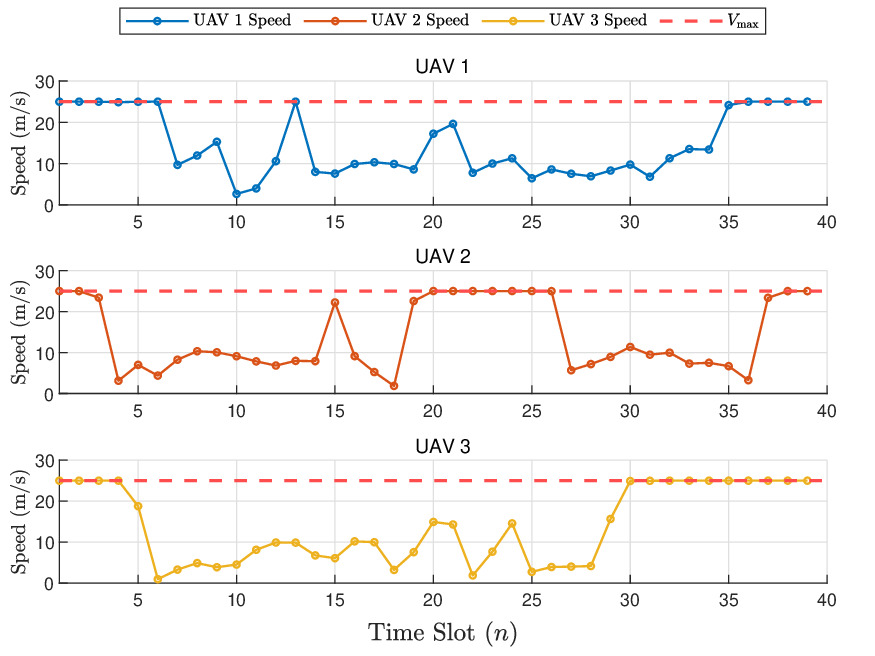}
    \caption{The speed of the UAVs in each time slot.}
    \label{fig:speed_per_slot}
\end{figure}

\begin{figure}[h]
    \centering
    \includegraphics[width=0.4\textwidth]{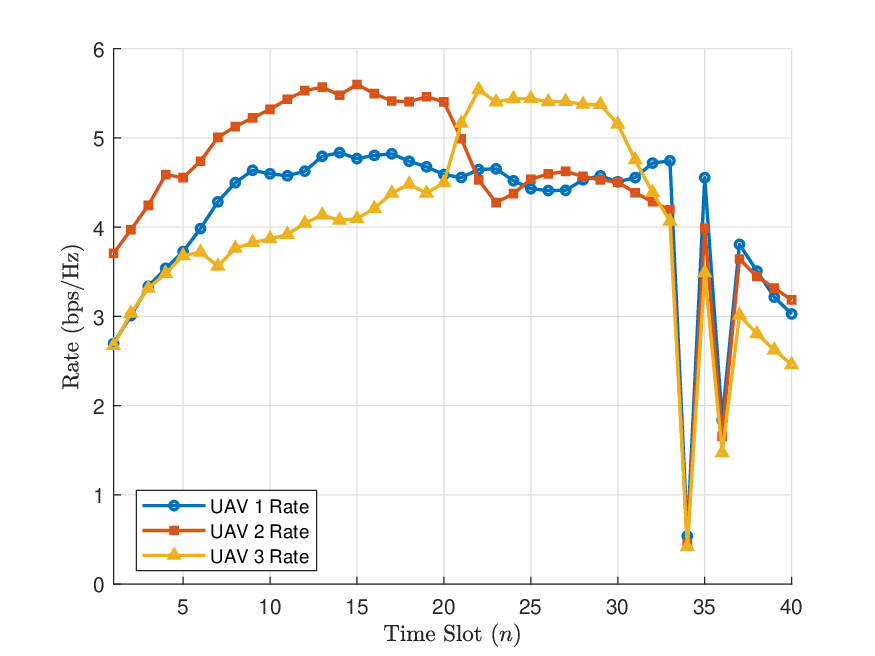}
    \caption{The communication rates of the UAVs in each time slot.}
    \label{fig:rate_per_slot}
\end{figure}

\begin{figure}[h]
    \centering
    \includegraphics[width=0.4\textwidth]{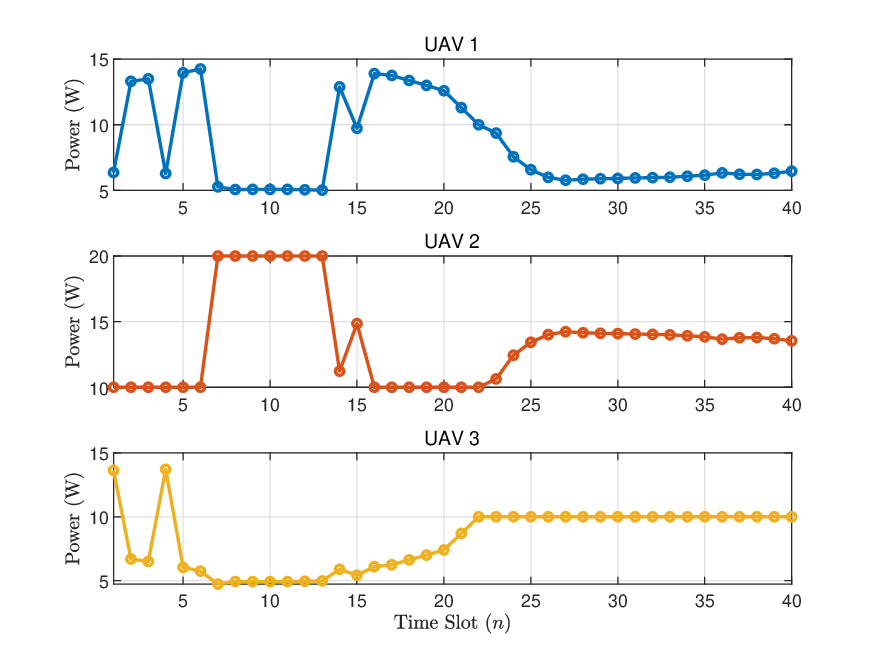}
    \caption{The power allocated by the BSs to each UAV in each time slot.}
    \label{fig:power_per_slot}
\end{figure}

For a more detailed analysis of the system's dynamic behavior, the slot-level results in Figs.~\ref{fig:speed_per_slot}--\ref{fig:power_per_slot} are presented for Position Setting 1 with the sensing requirement set to $R_{\min}^{\mathrm{MI}} = 60$\,bits, unless otherwise specified.

Fig.~\ref{fig:speed_per_slot} depicts the optimized flight speed profiles for each UAV over the mission duration. The speed profiles are consistent with the optimized trajectories. The UAVs fly fast when they are far from the BS cluster, and they slow down in the middle slots when they approach the BSs to increase dwell time under favorable channels. The maximum speed constraint $V_{\max}=25$\,m/s is always satisfied, which confirms the feasibility of the optimized trajectories.

Fig.~\ref{fig:rate_per_slot} presents the achieved communication rate at each time slot, which is inversely correlated with the flight speed profile shown previously. The communication rate is modest during the initial and final stages of the flight but increases substantially during the middle phase. This trend is consistent with the optimized trajectories, as the UAVs achieve higher channel gains when operating closer to the BSs in the middle of the mission. This strategic positioning directly translates to a higher SINR and a higher communication rate, demonstrating the effectiveness of the trajectory optimization in maximizing the primary communication objective.

Sharp valleys at slots 34 and 36 originate from dynamic time division. In these slots, the scheduler assigns a larger sensing fraction, which reduces the communication portion \(1-\delta[n]\). This choice helps the system satisfy the cumulative sensing constraint \(R_{\min}^{\mathrm{MI}}\) while preserving communication throughput in other slots. These results show that the proposed policy dynamically redistributes sensing time across slots according to the global sensing requirement and the instantaneous geometry, thereby balancing sensing and communication on a slot-by-slot basis.

Fig.~\ref{fig:power_per_slot} illustrates the dynamic communication power allocation strategy. The total power utilized by the system is not constant; the algorithm allocates significantly more power during the middle phase of the mission, approximately from time slots 10 to 30. This allocation strategy correlates directly with the proximity of UAVs to the BSs, as applying more power under favorable channel conditions maximizes the efficiency of rate enhancement. Moreover, power is distributed asymmetrically among the UAVs to prioritize the objective of sum communication rate maximization. The algorithm assigns a larger portion of the power budget to the UAV experiencing the best instantaneous channel conditions, a behavior exemplified by UAV 1 during its high-rate operational phase. This continuous adjustment of power distribution highlights the algorithm's adaptability to the rapidly changing mobile channel environment. The overall power expenditure profile mirrors the sum communication rate performance shown in Fig.~\ref{fig:rate_per_slot}, which confirms that power is a key resource strategically managed to increase communication throughput.

\subsection{Performance Comparison}

\begin{figure}[t]
    \centering
    \includegraphics[width=0.4\textwidth]{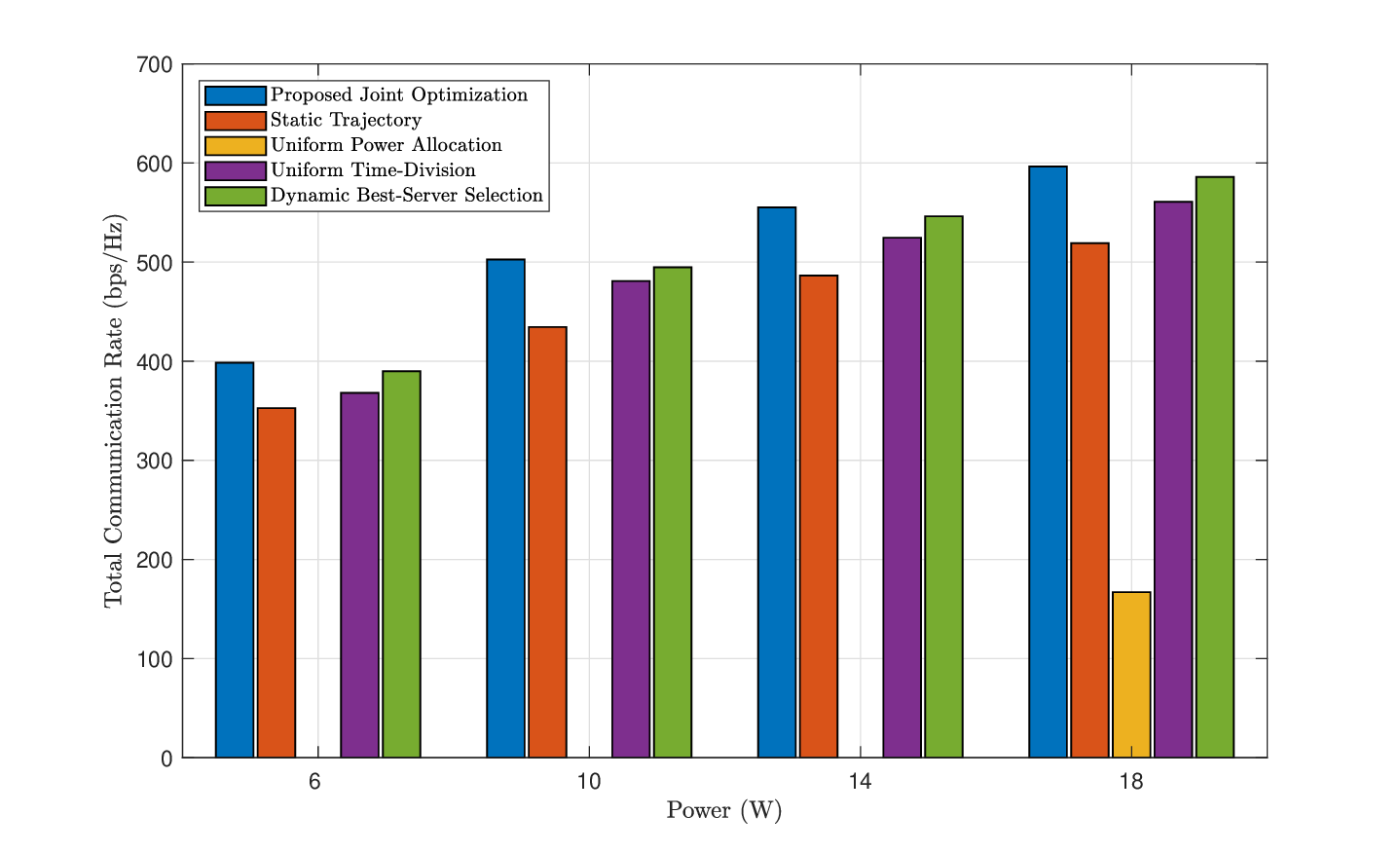}
    \caption{Sum communication rate of the proposed algorithm and benchmark schemes under various transmit powers.}
    \label{fig:rate_vs_power}
\end{figure}

Fig.~\ref{fig:rate_vs_power} reports the sum communication rate of the proposed scheme and four benchmarks under varying maximum transmit powers, with the sensing requirement fixed at \(R_{\min}^{\mathrm{MI}}=60\) bits. For all feasible schemes, the sum communication rate increases monotonically with transmit power, since a larger power budget improves the communication SINR and provides greater flexibility for balancing sensing and communication. The proposed algorithm consistently achieves the highest communication rate at all tested power levels, demonstrating the benefit of jointly optimizing UAV trajectory, power allocation, and dynamic time division.

Among the benchmarks, the ``Dynamic Best-Server Selection'' scheme performs better than the weaker ablation baselines in most power regimes, but remains consistently inferior to the proposed method. This indicates that although dynamic single-BS association with optimized resources is effective, cooperative multi-BS transmission provides an additional gain. The ``Uniform Time Division'' benchmark also outperforms the ``Static Trajectory'' benchmark, showing the importance of trajectory optimization. In contrast, the ``Uniform Power Allocation'' benchmark is often infeasible under low and moderate power budgets, since its fixed power policy cannot efficiently adapt communication and sensing resource usage. A feasible point appears only when \(P_{\max}\) is sufficiently high, for example \(18\) W, but its achieved rate is still much lower than those of the other schemes. Overall, these results confirm the importance of adaptive power control, slot-wise time allocation, trajectory optimization, and cooperative multi-BS serving under power-constrained sensing-communication trade-offs.

\begin{figure}[h]
    \centering
    \includegraphics[width=0.4\textwidth]{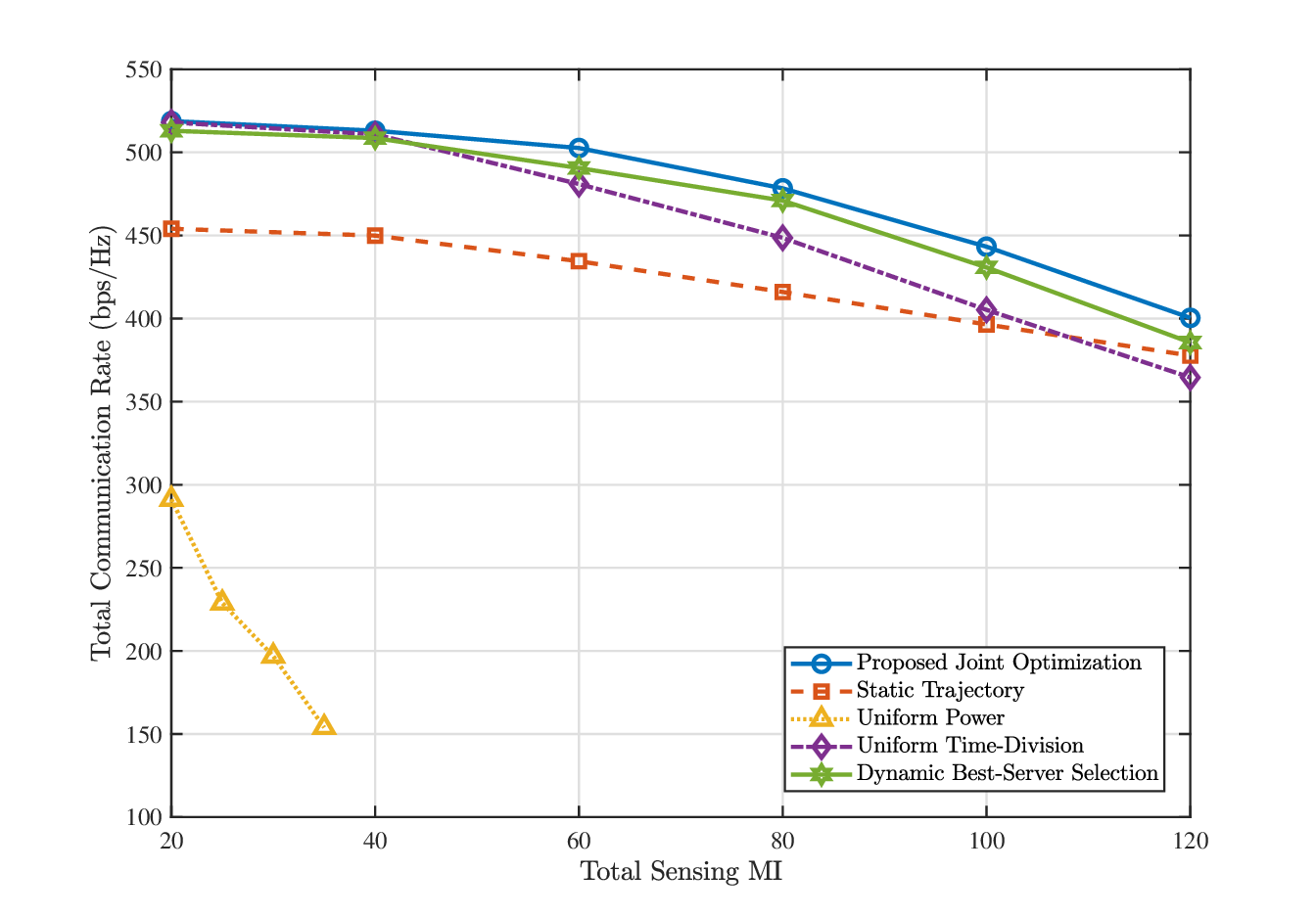}
    \caption{Sum communication rate of the proposed algorithm and benchmark schemes under various MI requirements.}
    \label{fig:rate_vs_MI}
\end{figure}

Fig.~\ref{fig:rate_vs_MI} illustrates the trade-off between the sum communication rate and the sensing requirement under \(P_{\max}=10\) W. For all schemes, the achievable sum communication rate decreases as the minimum MI requirement increases, since a stricter sensing target consumes more time and power resources and leaves fewer resources for communication. Across the entire tested range of \(R_{\min}^{\mathrm{MI}}\), the proposed scheme consistently achieves the highest sum communication rate, demonstrating its strongest ability to balance communication and sensing. Compared with the proposed scheme, the ``Static Trajectory'' benchmark is consistently inferior, confirming the importance of trajectory optimization in shaping favorable BS--UAV geometries. The ``Uniform Power Allocation'' benchmark performs worst and becomes infeasible under moderate or high MI requirements, highlighting the need for adaptive power control. The ``Uniform Time Division'' benchmark performs close to the proposed method when \(R_{\min}^{\mathrm{MI}}=20\) bits, because a nearly uniform sensing-communication split is sufficient in this low-demand regime, but degrades much faster as \(R_{\min}^{\mathrm{MI}}\) increases, revealing the value of slot-wise adaptive time allocation. The ``Dynamic Best-Server Selection'' benchmark, which restricts each UAV to a single serving BS in each slot while still optimizing trajectory, power allocation, and time division, outperforms the weaker ablation baselines but remains consistently below the proposed scheme. This gap shows that, beyond trajectory and resource optimization, cooperative multi-BS transmission brings an additional performance gain. Overall, the results confirm that the advantage of the proposed design comes from the joint optimization of trajectory, power allocation, and dynamic time division, together with cooperative multi-BS serving.

\subsection{Sensitivity Analysis}
\label{subsec:sensitivity_analysis}

To further reveal how key system parameters affect the optimized variables, we supplement a sensitivity analysis with respect to the MI threshold and the number of UAVs. Specifically, in the first group of experiments, the minimum cumulative sensing requirement is varied as $R_{\min}^{\mathrm{MI}} \in \{60,80,100,120\}$ while the number of UAVs is fixed as $K=3$. In the second group, the number of UAVs is varied as $K \in \{2,3,4\}$ while the sensing requirement is fixed as $R_{\min}^{\mathrm{MI}}=100$ bits. These experiments are intended to explicitly show how the MI threshold and the system scale influence the optimized time-division ratio, power allocation, and trajectory organization, rather than only reporting the final communication-rate improvement.

\begin{figure*}[t]
    \centering
    \includegraphics[width=0.75\textwidth]{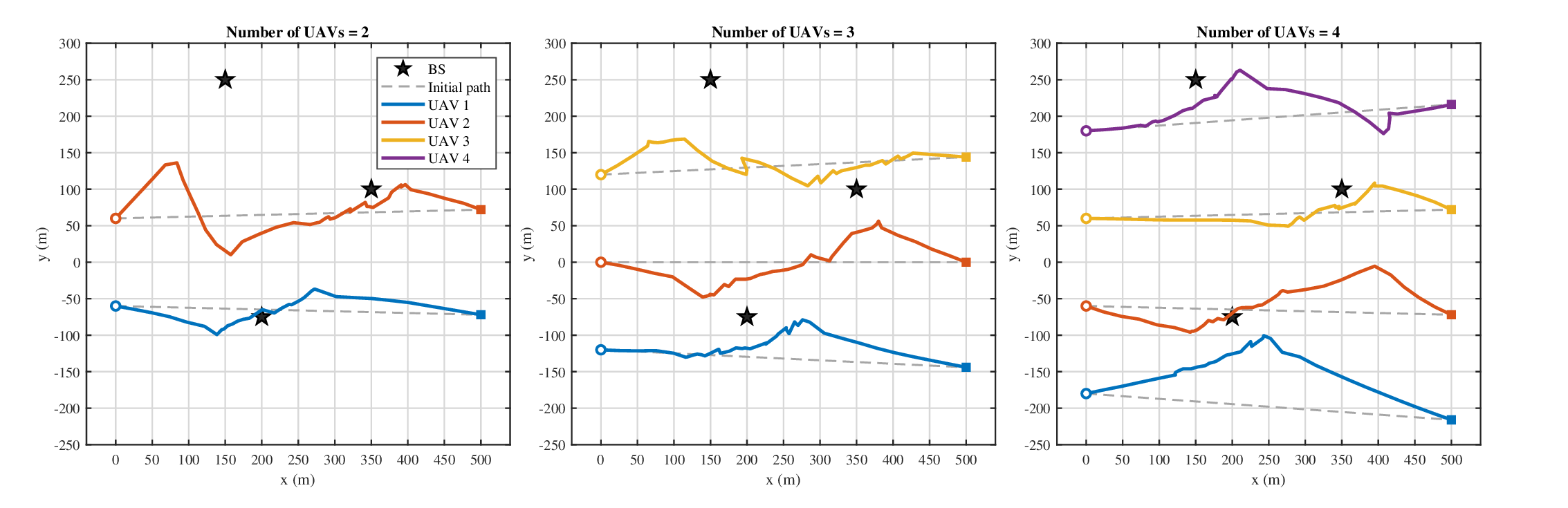}
    \caption{Optimized UAV trajectories under different numbers of UAVs.}
    \label{fig:traj_vs_k_sensitivity}
\end{figure*}

\begin{figure*}[t]
    \centering
    \begin{subfigure}{\textwidth}
        \centering
        \includegraphics[width=0.90\textwidth]{Figure/MI_1x4_boxedlegend.png}
        \caption{Sensitivity to the MI threshold: communication-rate variation, time-division summary, power-allocation variation, and slot-wise time-division pattern.}
        \label{fig:summary_vs_mi_sensitivity}
    \end{subfigure}

    \begin{subfigure}{\textwidth}
        \centering
        \includegraphics[width=0.90\textwidth]{Figure/K_1x4_boxedlegend.png}
        \caption{Sensitivity to the number of UAVs: communication-rate variation, time-division summary, spatial separation, and power-allocation variation.}
        \label{fig:summary_vs_k_sensitivity}
    \end{subfigure}

    \caption{Sensitivity of optimized non-trajectory variables to the MI threshold and the number of UAVs.}
    \label{fig:nontraj_sensitivity_combined}
\end{figure*}

Fig.~\ref{fig:traj_vs_k_sensitivity} shows how the UAV trajectories change with system scale, while Fig.~\ref{fig:nontraj_sensitivity_combined} summarizes the sensitivity of the optimized non-trajectory variables to the MI threshold and the number of UAVs. Together, they illustrate how the proposed joint design adapts time allocation, power allocation, and spatial organization under different conditions.

We first examine the impact of the MI threshold. From Fig.~\ref{fig:nontraj_sensitivity_combined}(a), the total communication rate decreases as $R_{\min}^{\mathrm{MI}}$ increases, since a stricter sensing requirement consumes more resources. Meanwhile, the optimized time-division factor increases. In particular, the mean value of $\delta[n]$ rises with the MI threshold, while the maximum value quickly reaches its upper bound, indicating that some slots become sensing-dominant under stringent MI requirements. The slot-wise heatmap further shows that this adjustment is concentrated in a subset of critical slots rather than applied uniformly over time, highlighting the temporal selectivity of the proposed dynamic time-division design.

The power-allocation results in Fig.~\ref{fig:nontraj_sensitivity_combined}(a) further show that the response to a larger MI threshold is achieved through coupled adjustments in time division, trajectory, and power allocation. As a result, the communication and sensing powers do not necessarily vary monotonically with $R_{\min}^{\mathrm{MI}}$, which is consistent with the coupled structure of the optimization problem.

We next examine the impact of the number of UAVs. As shown in Fig.~\ref{fig:nontraj_sensitivity_combined}(b), the total communication rate increases with $K$, indicating that adding more UAVs enlarges the overall service demand and provides additional spatial flexibility for cooperative transmission. At the same time, the average time-division factor decreases sharply from $K=2$ to $K=3$ and then remains relatively low at $K=4$. This trend is closely related to the fixed global MI threshold ($R_{\min}^{\mathrm{MI}} = 100$ bits). Under this setting, increasing the number of UAVs can accelerate the accumulation of sensing information in the optimized solution, so the same cumulative MI target can be satisfied with a smaller average sensing-time fraction.

The spatial-separation and power-allocation results in Fig.~\ref{fig:nontraj_sensitivity_combined}(b) provide further insight. The mean pairwise distance increases with $K$, showing that the fleet spreads over a wider region as more UAVs are introduced. In contrast, the minimum pairwise distance decreases, indicating that local proximity may still occur in some slots even though the overall fleet expands. This trend is also reflected in Fig.~\ref{fig:traj_vs_k_sensitivity}, where the trajectories become more differentiated as $K$ increases. Moreover, although the total average communication and sensing powers both increase with $K$, the average BS power allocated per UAV decreases. This indicates that a larger fleet improves the overall system capability while reducing the average per-UAV resource burden on the infrastructure.

Overall, the MI threshold mainly increases sensing-resource pressure, whereas the number of UAVs under a fixed global MI target accelerates the accumulation of sensing information and strengthens spatial task sharing. The sensitivity results therefore complement the performance comparisons by showing how key parameters affect time division, power allocation, and trajectory organization.

\section{Conclusion}

In this paper, we studied a cooperative UAV-assisted ISAC framework for the low-altitude economy (LAE), where each time slot consists of a sensing phase followed by a communication phase. We formulated a mission-level sum-rate maximization problem under a cumulative sensing MI constraint and practical UAV mobility constraints, and developed an AO--SCA algorithm to jointly optimize UAV trajectories, communication and sensing power allocation, and the dynamic time-division ratio. Numerical results showed that the proposed design consistently outperforms the benchmark schemes and effectively balances the sensing-communication trade-off. In addition, the complexity discussion and sensitivity analysis provided further insight into how the MI threshold and the UAV swarm size affect the optimized spatial and temporal resource allocation. These results demonstrate the effectiveness and practical potential of the proposed framework for cooperative ISAC in dynamic air-to-ground networks.

\bibliographystyle{ieeetr}
\bibliography{references}

@ARTICLE{10906066,
author={Tang, Jun and Yu, Yiming and Pan, Cunhua and Ren, Hong and Wang, Dongming and Wang, Jiangzhou and You, Xiaohu},
journal={IEEE Trans. Wireless Commun.},
title={Cooperative {ISAC}-Empowered Low-Altitude Economy},
year={2025},
volume={24},
number={5},
pages={3837-3853},
doi={10.1109/TWC.2025.3542399},
ISSN={1558-2248},
month={May}
}

@article{li2026cooperative,
author={Li, Fangzhi and Ren, Zhichu and Pan, Cunhua and Ren, Hong and Jin, Jing and Wang, Qixing and Wang, Jiangzhou},
journal={Sci. China Inf. Sci.},
title={Cooperative sensing and communication beamforming design for low-altitude economy},
year={2026},
volume={69},
number={2},
pages={122304},
doi={10.1007/s11432-025-4640-y},
month={Feb.}
}

@article{tradeoff_isac_1,
author={Keskin, Musa Furkan and Mojahedian, Mohammad Mahdi and Lacruz, Jesus O. and Marcus, Carina and Eriksson, Olof and Giorgetti, Andrea and Widmer, Joerg and Wymeersch, Henk},
journal={IEEE Trans. Wireless Commun.},
title={Fundamental Trade-Offs in Monostatic {ISAC}: A Holistic Investigation Toward {6G}},
year={2025},
volume={24},
number={9},
pages={7856-7873},
doi={10.1109/TWC.2025.3563197},
ISSN={1558-2248},
month={Sep.}
}

@article{tradeoff_isac_2,
author={Liu, Fan and Masouros, Christos and Petropulu, Athina P. and Griffiths, Hugh and Hanzo, Lajos},
journal={IEEE Trans. Commun.},
title={Joint Radar and Communication Design: Applications, State-of-the-Art, and the Road Ahead},
year={2020},
volume={68},
number={6},
pages={3834-3862},
doi={10.1109/TCOMM.2020.2973976},
ISSN={1558-0857},
month={Jun.}
}

@article{survey_isac_1,
author={Liu, Fan and Cui, Yuanhao and Masouros, Christos and Xu, Jie and Han, Tony Xiao and Eldar, Yonina C. and Buzzi, Stefano},
journal={IEEE J. Sel. Areas Commun.},
title={Integrated Sensing and Communications: Toward Dual-Functional Wireless Networks for {6G} and Beyond},
year={2022},
volume={40},
number={6},
pages={1728-1767},
doi={10.1109/JSAC.2022.3156632},
ISSN={1558-0008},
month={Mar.}
}

@article{survey_isac_2,
author={Wen, Dingzhu and Zhou, Yong and Li, Xiaoyang and Shi, Yuanming and Huang, Kaibin and Letaief, Khaled B.},
journal={IEEE Commun. Surv. Tutorials},
title={A Survey on Integrated Sensing, Communication, and Computation},
year={2024},
volume={},
number={},
pages={1-1},
doi={10.1109/COMST.2024.3521498},
ISSN={1553-877X},
month={Dec.}
}

@INPROCEEDINGS{survey_isac_3,
author={Pin Tan, Danny Kai and He, Jia and Li, Yanchun and Bayesteh, Alireza and Chen, Yan and Zhu, Peiying and Tong, Wen},
booktitle={Proc. IEEE Int. Online Symp. on Joint Commun. \& Sens. (JC\&S)},
title={Integrated Sensing and Communication in {6G}: Motivations, Use Cases, Requirements, Challenges and Future Directions},
year={2021},
pages={1-6},
doi={10.1109/JCS52304.2021.9376324},
month={Feb.}
}

@article{uav_comm_survey_1,
author={Zeng, Yong and Zhang, Rui and Lim, Teng Joon},
journal={IEEE Commun. Mag.},
title={Wireless Communications with Unmanned Aerial Vehicles: Opportunities and Challenges},
year={2016},
volume={54},
number={5},
pages={36-42},
doi={10.1109/MCOM.2016.7470933},
ISSN={1558-1896},
month={May}
}

@article{uav_comm_survey_2,
author={Mozaffari, Mohammad and Saad, Walid and Bennis, Mehdi and Nam, Young-Han and Debbah, Mérouane},
journal={IEEE Commun. Surv. Tutorials},
title={A Tutorial on {UAVs} for Wireless Networks: Applications, Challenges, and Open Problems},
year={2019},
volume={21},
number={3},
pages={2334-2360},
doi={10.1109/COMST.2019.2902862},
ISSN={1553-877X},
month={3rd Quart.}
}

@article{uav_isac_survey_1,
author={Mu, Junsheng and Zhang, Ronghui and Cui, Yuanhao and Gao, Ning and Jing, Xiaojun},
journal={IEEE Commun. Mag.},
title={{UAV} Meets Integrated Sensing and Communication: Challenges and Future Directions},
year={2023},
volume={61},
number={5},
pages={62-67},
doi={10.1109/MCOM.008.2200510},
ISSN={1558-1896},
month={May}
}

@INPROCEEDINGS{uav_isac_survey_2,
author={Zhang, Kexin and Shen, Chao},
booktitle={Proc. IEEE Veh. Technol. Conf. (VTC-Fall)},
title={{UAV} Aided Integrated Sensing and Communications},
year={2021},
pages={1-6},
doi={10.1109/VTC2021-Fall52928.2021.9625578},
month={Sep.}
}

@article{uav_isac_joint_1,
author={Liu, Zechen and Liu, Xin and Liu, Yuemin and Leung, Victor C. M. and Durrani, Tariq S.},
journal={IEEE Trans. Wireless Commun.},
title={{UAV} Assisted Integrated Sensing and Communications for Internet of Things: {3D} Trajectory Optimization and Resource Allocation},
year={2024},
volume={23},
number={8},
pages={8654-8667},
doi={10.1109/TWC.2024.3352985},
ISSN={1558-2248},
month={Aug.}
}

@INPROCEEDINGS{uav_isac_joint_2,
author={Lyu, Zhonghao and Zhu, Guangxu and Xu, Jie},
booktitle={Proc. IEEE Int. Conf. Commun. (ICC)},
title={Joint Trajectory and Beamforming Design for {UAV}-Enabled Integrated Sensing and Communication},
year={2022},
pages={1593-1598},
doi={10.1109/ICC45855.2022.9839031},
month={May}
}

@article{uav_isac_joint_3,
author={Zhang, Xian and Peng, Mugen and Liu, Chenxi},
journal={IEEE Trans. Veh. Technol.},
title={Sensing-Assisted Beamforming and Trajectory Design for {UAV}-Enabled Networks},
year={2024},
volume={73},
number={3},
pages={3804-3819},
doi={10.1109/TVT.2023.3326407},
ISSN={1939-9359},
month={Oct.}
}

@article{jing2024isac,
author={Jing, Xiaoye and Liu, Fan and Masouros, Christos and Zeng, Yong},
journal={IEEE Trans. Wireless Commun.},
title={{ISAC} From the Sky: {UAV} Trajectory Design for Joint Communication and Target Localization},
year={2024},
volume={23},
number={10},
pages={12857-12872},
doi={10.1109/TWC.2024.3396571},
ISSN={1558-2248},
month={Oct.}
}

@article{resource_alloc_isac_1,
author={Liu, An and Huang, Zhe and Li, Min and Wan, Yubo and Li, Wenrui and Han, Tony Xiao and Liu, Chenchen and Du, Rui and Tan, Danny Kai Pin and Lu, Jianmin and Shen, Yuan and Colone, Fabiola and Chetty, Kevin},
journal={IEEE Commun. Surv. Tutorials},
title={A Survey on Fundamental Limits of Integrated Sensing and Communication},
year={2022},
volume={24},
number={2},
pages={994-1034},
doi={10.1109/COMST.2022.3149272},
ISSN={1553-877X},
month={2nd Quart.}
}

@article{resource_alloc_isac_2,
author={Lyu, Zhonghao and Zhu, Guangxu and Xu, Jie},
journal={IEEE Trans. Wireless Commun.},
title={Joint Maneuver and Beamforming Design for {UAV}-Enabled Integrated Sensing and Communication},
year={2023},
volume={22},
number={4},
pages={2424-2440},
doi={10.1109/TWC.2022.3211533},
ISSN={1558-2248},
month={Apr.}
}

@ARTICLE{9858656,
author={Meng, Kaitao and Wu, Qingqing and Ma, Shaodan and Chen, Wen and Wang, Kunlun and Li, Jun},
journal={IEEE Trans. Wireless Commun.},
title={Throughput Maximization for {UAV}-Enabled Integrated Periodic Sensing and Communication},
year={2023},
volume={22},
number={1},
pages={671-687},
doi={10.1109/TWC.2022.3197623},
ISSN={1558-2248},
month={Aug.}
}

@article{meng2022uav,
author={Meng, Kaitao and Wu, Qingqing and Ma, Shaodan and Chen, Wen and Quek, Tony Q. S.},
journal={IEEE Wireless Commun. Lett.},
title={{UAV} Trajectory and Beamforming Optimization for Integrated Periodic Sensing and Communication},
year={2022},
volume={11},
number={6},
pages={1211-1215},
doi={10.1109/LWC.2022.3161338},
ISSN={2162-2345},
month={Jun.}
}

@article{khalili2024efficient,
author={Khalili, Ata and Rezaei, Atefeh and Xu, Dongfang and Dressler, Falko and Schober, Robert},
journal={IEEE Trans. Wireless Commun.},
title={Efficient {UAV} Hovering, Resource Allocation, and Trajectory Design for {ISAC} With Limited Backhaul Capacity},
year={2024},
volume={23},
number={11},
pages={17635-17650},
doi={10.1109/TWC.2024.3455370},
ISSN={1558-2248},
month={Nov.}
}

@article{pan2023cooperative,
author={Pan, Yu and Li, Ruoguang and Da, Xinyu and Hu, Hang and Zhang, Miao and Zhai, Dong and Cumanan, Kanapathippillai and Dobre, Octavia A.},
journal={IEEE Trans. Veh. Technol.},
title={Cooperative Trajectory Planning and Resource Allocation for {UAV}-Enabled Integrated Sensing and Communication Systems},
year={2024},
volume={73},
number={5},
pages={6502-6516},
doi={10.1109/TVT.2023.3337106},
ISSN={1939-9359},
month={May}
}

@inproceedings{cheng2024networked,
author={Cheng, Gaoyuan and Song, Xianxin and Lyu, Zhonghao and Xu, Jie},
booktitle={Proc. IEEE/CIC International Conference on Communications in China (ICCC)},
title={Networked {ISAC} for Low-Altitude Economy: Transmit Beamforming and {UAV} Trajectory Design},
year={2024},
pages={78-83},
doi={10.1109/ICCC62479.2024.10681882},
month={Aug.}
}

@article{cheng2025networked,
author={Cheng, Gaoyuan and Song, Xianxin and Lyu, Zhonghao and Xu, Jie},
journal={IEEE Trans. Commun.},
title={Networked {ISAC} for Low-Altitude Economy: Coordinated Transmit Beamforming and {UAV} Trajectory Design},
year={2025},
volume={},
number={},
pages={1-1},
doi={10.1109/TCOMM.2025.3541027},
ISSN={1558-0857},
month={Feb.}
}

@article{wang2024isac,
author={Wang, Yi and Zu, Keke and Xiang, Luping and Zhang, Qixun and Feng, Zhiyong and Hu, Jie and Yang, Kun},
journal={IEEE Trans. Wireless Commun.},
title={ISAC Enabled Cooperative Detection for Cellular-Connected {UAV} Network},
year={2025},
volume={24},
number={2},
pages={1541-1554},
doi={10.1109/TWC.2024.3509978},
ISSN={1558-2248},
month={Feb.}
}

@article{zhang2024joint,
author={Zhang, Ruizhi and Zhang, Ying and Tang, Rui and Zhao, Huapeng and Xiao, Qing and Wang, Chenye},
journal={IEEE Internet Things J.},
title={A Joint {UAV} Trajectory, User Association, and Beamforming Design Strategy for Multi-{UAV}-Assisted {ISAC} Systems},
year={2024},
volume={11},
number={18},
pages={29360-29374},
doi={10.1109/JIOT.2024.3430390},
ISSN={2327-4662},
month={Sep.}
}

@article{pang2024dynamic,
author={Pang, Xiaowei and Guo, Shaoyong and Tang, Jie and Zhao, Nan and Al-Dhahir, Naofal},
journal={IEEE Trans. Wireless Commun.},
title={Dynamic {ISAC} Beamforming Design for {UAV}-Enabled Vehicular Networks},
year={2024},
volume={23},
number={11},
pages={16852-16864},
doi={10.1109/TWC.2024.3447779},
ISSN={1558-2248},
month={Nov.}
}

@article{chai2024precoding,
author={Chai, Rong and Cui, Xianglin and Sun, Ruijin and Zhao, Dongmei and Chen, Qianbin},
journal={IEEE Trans. Veh. Technol.},
title={Precoding and Trajectory Design for {UAV}-Assisted Integrated Communication and Sensing Systems},
year={2024},
volume={73},
number={9},
pages={13151-13163},
doi={10.1109/TVT.2024.3390693},
ISSN={1939-9359},
month={Sep.}
}

@inproceedings{gao2024trajectory,
author={Gao, Qian and Zhong, Ruikang and Liu, Yuanwei},
booktitle={Proc. IEEE Global Commun. Conf. (GLOBECOM)},
title={Trajectory and Beamforming Optimization in {UAV}-enabled {ISAC} System},
year={2024},
pages={1527-1532},
doi={10.1109/GLOBECOM52923.2024.10901014},
month={Dec.}
}

@article{zhou2024temporal,
author={Zhou, Shengcai and Yang, Halvin and Xiang, Luping and Yang, Kun},
journal={IEEE Trans. Commun.},
title={Temporal-Assisted Beamforming and Trajectory Prediction in Sensing-Enabled {UAV} Communications},
year={2024},
volume={},
number={},
pages={1-1},
doi={10.1109/TCOMM.2024.3519546},
ISSN={1558-0857},
month={Dec.}
}

@article{sca_tutorial,
author={Sun, Ying and Babu, Prabhu and Palomar, Daniel P.},
journal={IEEE Trans. Signal Process.},
title={Majorization–Minimization Algorithms in Signal Processing, Communications, and Machine Learning},
year={2017},
volume={65},
number={3},
pages={794-816},
doi={10.1109/TSP.2016.2601299},
ISSN={1941-0476},
month={Feb.}
}

@article{sensing_metric_mi,
author={Bazzi, Ahmad and Chafii, Marwa},
journal={IEEE Trans. Commun.},
title={Mutual Information Based Pilot Design for {ISAC}},
year={2025},
volume={73},
number={9},
pages={7914-7930},
doi={10.1109/TCOMM.2025.3545658},
ISSN={1558-0857},
month={Feb.}
}

@article{sensing_metric_crb,
author={Stoica, P. and Nehorai, A.},
journal={IEEE Trans. Acoust., Speech, Signal Process.},
title={Performance Study of Conditional and Unconditional Direction-of-Arrival Estimation},
year={1990},
volume={38},
number={10},
pages={1783-1795},
doi={10.1109/29.60109},
ISSN={0096-3518},
month={Oct.}
}

@article{sensing_metric_tradeoff,
author={Chiriyath, Alex R. and Paul, Bryan and Bliss, Daniel W.},
journal={IEEE Trans. Cogn. Commun. Netw.},
title={Radar–Communications Convergence: Coexistence, Cooperation, and Co-Design},
year={2017},
volume={3},
number={1},
pages={1-12},
doi={10.1109/TCCN.2017.2666266},
ISSN={2332-7731},
month={Mar.}
}

@article{uav_traj_comm_1,
author={Wu, Qingqing and Zeng, Yong and Zhang, Rui},
journal={IEEE Trans. Wireless Commun.},
title={Joint Trajectory and Communication Design for Multi-{UAV} Enabled Wireless Networks},
year={2018},
volume={17},
number={3},
pages={2109-2121},
doi={10.1109/TWC.2017.2789293},
ISSN={1558-2248},
month={Mar.}
}

\end{document}